\let\Xdocument\document
\documentclass{iau}
\let\document\Xdocument

\usepackage{amsmath}
\usepackage{graphicx}
\usepackage{multirow}
\begin{document}

\lefttitle{Houde et al.}
\righttitle{Variability, flaring and coherence -- the maser and superradiance regimes}

\jnlPage{1}{7}
\jnlDoiYr{2023}
\doival{TBD}
\aopheadtitle{Proceedings IAU Symposium}
\editors{T. Hirota,  H. Imai, K. Menten, \& Y. Pihlstr\"om, eds.}

\title{Variability, flaring and coherence -- the complementarity of the maser and superradiance regimes}

\author{Martin Houde$^{1}$, 
             Fereshteh Rajabi$^{2}$,
             Gordon C. MacLeod$^{3,4}$,
             Sharmila Goedhart$^{5,6}$,
             Yoshihiro Tanabe$^{7}$,
             Stefanus P. van den Heever$^{4}$,
             Christopher M. Wyenberg$^{8}$, 
             and Yoshinori Yonekura$^{7}$}
\affiliation{$^{1}$The University of Western Ontario, London, Ontario N6A 3K7, Canada \\
                 $^{2}$McMaster University, Hamilton, Ontario L8S 4M1, Canada\\
                 $^{3}$The Open University of Tanzania, P.O. Box 23409, Dar-Es-Salaam, Tanzania \\
                 $^{4}$Hartebeesthoek Radio Astronomy Observatory, P.O. Box 443, Krugersdorp, 1741, South Africa \\
                 $^{5}$South African Radio Astronomy Observatory, 2 Fir Street, Black River Park, Observatory 7925, South Africa \\
                 $^{6}$Center for Space Research, North-West University, Potchefstroom Campus, Private Bag X6001, Potchefstroom 2520, South Africa \\
                 $^{7}$Center for Astronomy, Ibaraki University, 2-1-1 Bunkyo, Mito, Ibaraki 310-8512, Japan \\
                 $^{8}$Institute for Quantum Computing and Department of Physics and Astronomy, The University of Waterloo, 200 University Ave. West, Waterloo, Ontario N2L 3G1, Canada}

\begin{abstract}
We discuss the role that coherence phenomena can have on the intensity variability of spectral lines associated with maser radiation. We do so by introducing the fundamental cooperative radiation phenomenon of (Dicke’s) superradiance and discuss its complementary nature to the maser action, as well as its role in the flaring behaviour of some maser sources. We will consider examples of observational diagnostics that can help discriminate between the two, and identify superradiance as the source of the latter. More precisely, we show how superradiance readily accounts for the different time-scales observed in the multi-wavelength monitoring of the periodic flaring in G9.62+0.20E.
\end{abstract}


\maketitle

\section{Introduction\label{sec:Introduction}}

The monitoring of astronomical objects harboring sources of maser radiation reveals ubiquitous variability in the measured intensities over a range of time-scales and light curve patterns. Most intriguing are strong flaring behaviours, which can consist of isolated or recurring events. Periodic flaring sources are of particular interest because of, among other things, their potential for shedding light on the nature of the engines at the root of the observed periodicities. Accordingly, a growing number of such objects have been discovered and closely monitored with several spectral transitions in the recent past.         

\begin{figure}
    \begin{center}
    \includegraphics[trim={0 10cm 0. 0.},clip,width=0.8\textwidth]{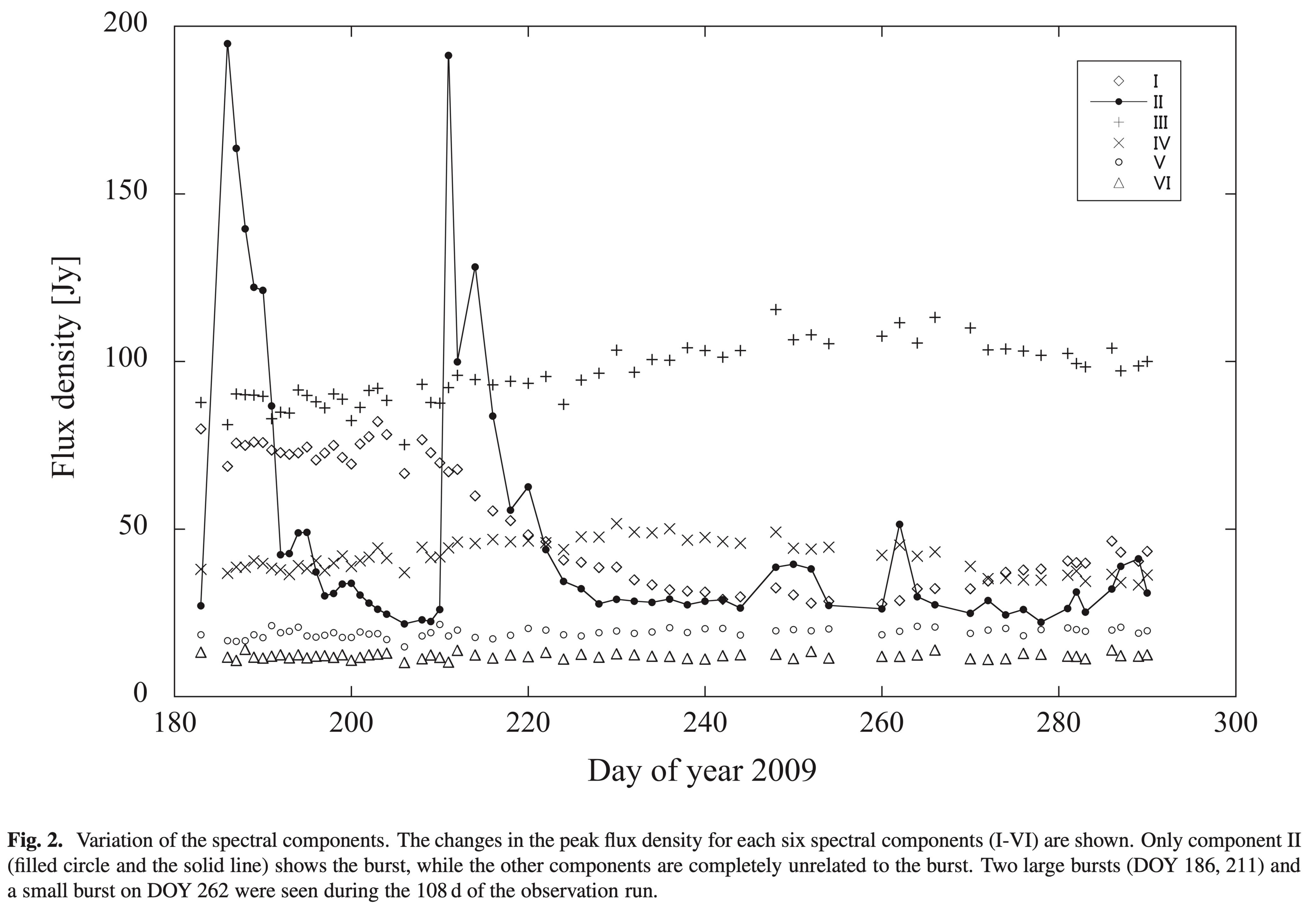}
    \caption{Flaring in the methanol 6.7~GHz transition for the G33.64--0.21 high mass star-forming region. Six different masing spectral features are detected in this line, but only one (Feature II) shows two strong flares exhibiting exceedingly fast rise times. Taken from \citet{Fujisawa2012}.}
    \label{fig:G33-Fujisawa}
    \end{center}
\end{figure}

Recurring and periodic flaring sources also often display behaviours that can seem difficult to explain theoretically. Examples can involve one spectral spectral transition (observed at different systemic velocities) or the comparison between several lines. Figure \ref{fig:G33-Fujisawa} shows the case of a recurring flare observed by \citet{Fujisawa2012} in the methanol 6.7~GHz transition for the G33.64--0.21 high mass star-forming region. Although six different masing spectral features are detected in this line, only one (Feature II) shows two strong flares with an amplification of a factor of approximately eight relative to the quiescent flux level and exceedingly fast rise times. The challenge in making sense of these observations, besides the nature of the flares themselves and their source, resides in explaining why features from the same spectral transitions but at different velocities could exhibit such vastly different responses to a common excitation.  

\begin{figure}
    \begin{center}
    \includegraphics[trim={0 8cm 0. 0.},clip,width=0.8\textwidth]{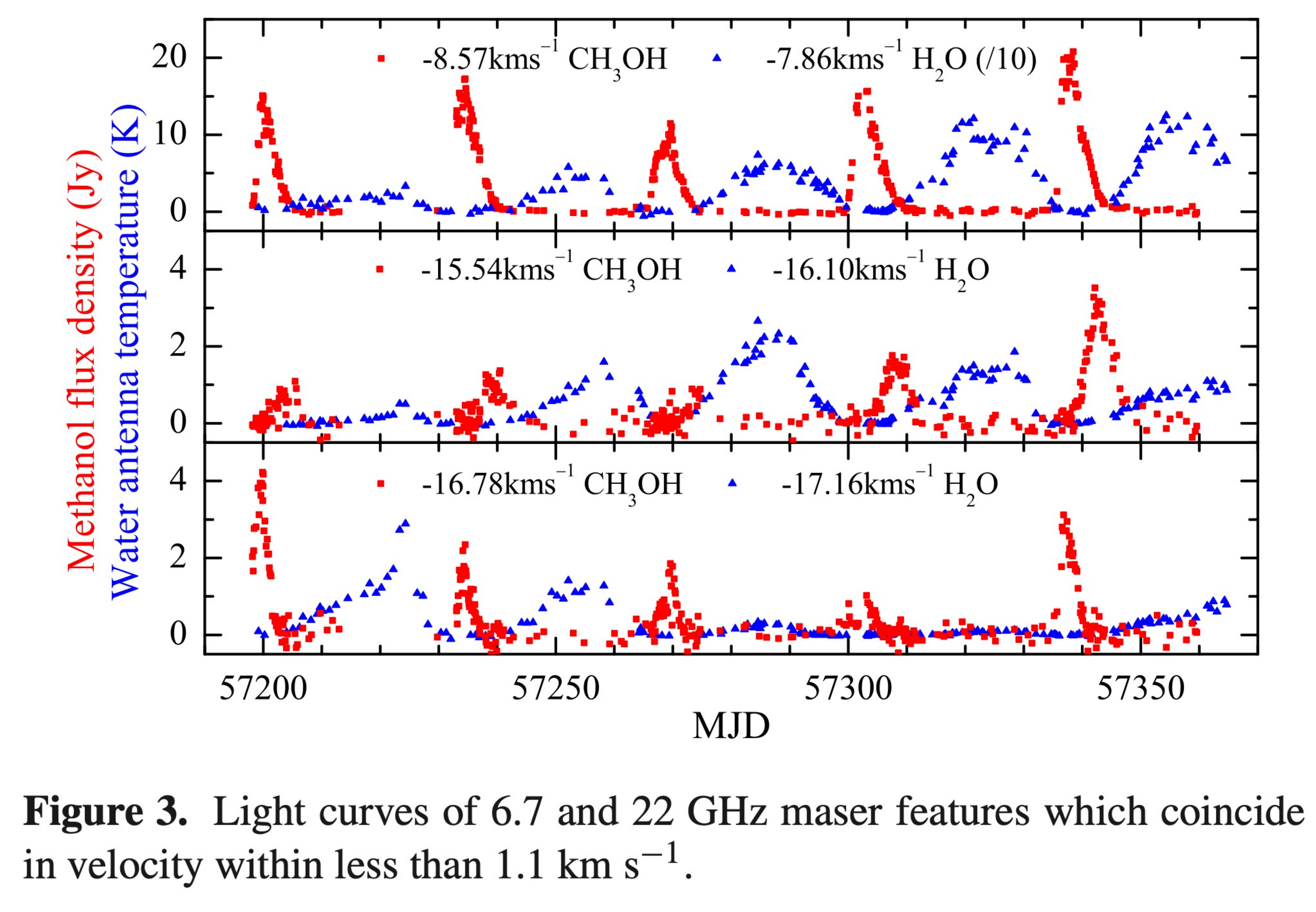}
    \caption{Flaring in the methanol 6.7~GHz and water 22~GHz transitions for the G107.298+5.639 star-forming region. The methanol 6.7~GHz and water 22~GHz flares are seen to alternate with the flux density of one transition peaking when the other reaches a minimum. Also to be noted are the different time-scales of the flares, where the methanol 6.7~GHz flares are consistently of shorter duration than the water 22~GHz features. Taken from \citet{Szymczak2016}.}
    \label{fig:G107-Szymczak}
    \end{center}
\end{figure}

Figure \ref{fig:G107-Szymczak} shows observations obtained by \citet{Szymczak2016} in methanol 6.7~GHz and water 22~GHz during a monitoring campaign of the G107.298+5.639 star-forming region (see also \citealt{Olech2020}). This source exhibits flaring in multiple spectral transitions at a common period of 34.4~d. More interestingly, and as shown in the figure, the methanol 6.7~GHz and water 22~GHz flares are seen to alternate with the flux density of one transition peaking when the other reaches a minimum. Also to be noted are the different time-scales of the flares, where the methanol 6.7~GHz flares are consistently of shorter duration than the water 22~GHz features. While the alternation of the flares between the two species can perhaps be explained by the effect of an infrared (methanol-) pumping source on the dust temperature and its quenching effect on the water 22~GHz population inversion \citep{Szymczak2016}, their different time-scales are more problematic.  

There are also more fundamental questions pertaining to the physics underlying the flaring phenomenon. That is, when subjected to an excitation signal (e.g., the infrared pump source or a seed electric field signal for a maser-hosting region) a physical system will often exhibit a transient response before settling into a steady-state regime. It has long been established in the quantum optics community that for an ensemble of radiators (i.e., atoms or molecules) exhibiting a population inversion the two regimes are associated with different radiation processes. That is, stimulated emission rules the quasi steady-state regime while Dicke's superradiance is at play during the transient phase \citep{Dicke1954,Feld1980}. One may wonder whether this dichotomy also applies to astrophysical sources and could be essential for explaining some of the characteristics of flares emanating from maser-hosting sources.   

In this review, we will use observations from multi-transition monitoring of the G9.62+0.20E high mass star-forming region to probe the physics underlying the periodic flaring observed in this source and determine whether different processes characterize the bursting and quiescent phases. The paper is structure as follows. We first introduce Dicke's superradiance and its study at the laboratory level in Sec. \ref{sec:superradiance}, while we discuss the association of the transient and quasi steady-state regimes to the superradiance and maser phenomena, respectively, as well as their manifestation as two independent limits of the Maxwell-Bloch equations in Sec. \ref{sec:complementarity}. Finally, in Sec. \ref{sec:G9.62} we model existing multi-transition data obtained for G9.62+0.20E and show that superradiance can naturally account for the diversity of time-scales observed in this source. 

\section{Dicke's superradiance\label{sec:superradiance}}

Given that masers were first discovered in the laboratory \citep{Gordon1954,Gordon1955}, it may be appropriate to inquire whether scientists in the quantum optics community have studied or have performed experiments on ``flaring'' with such systems? The answer to this question is a resounding ``yes.'' This is due to the introduction of the superradiance phenomenon by \citet{Dicke1954}, which, it is interesting to note, actually predates that of the maser. In his foundational paper, Dicke pointed out that molecules in a gas (atoms could also be used but we focus on molecules as we will be dealing with corresponding transitions in this review) may not always radiate independently, whereas their interaction with a common electromagnetic field will bring a quantum mechanical entanglement between them. Dicke thus treated the ensemble of molecules as a ``single quantum mechanical system'' and proceeded in calculating spontaneous emission rate for the gas when focusing on a single spectral transition. 

Under conditions where velocity coherence and population inversion prevail, he showed that the commonly assumed independent spontaneous emission process is replaced by a transient phenomenon he termed ``superradiance.'' With superradiance radiation from the ensemble of molecules happens on a photon cascade. For example, when the $N\gg 1$ molecules are all in the excited state the emission of photons takes place at transition rates ranging from $N\Gamma_0$ (at the beginning and the end of the cascade) to $\sim N^2\Gamma_0/4$ (in the middle of the cascade), with $\Gamma_0$ the single-molecule transition rate (i.e., the spontaneous emission rate of the transition). This results in powerful bursts of radiation intensity lasting on a time-scale on the order of $T_\mathrm{R}\propto 1/\left(N\Gamma_0\right)$\footnote{We will soon give a more precise definition of the superradiance time-scale.}, while the coherence nature of the radiation and the process can be attested through the scaling of the peak intensity with $N^2$. Evidently, coherence in the radiation cannot occur in all directions whenever the molecules are separated by a finite distance. In this case, Dicke showed that superradiance happens only in well-defined radiation modes with successive photons exhibiting increasing correlation in their propagation direction, and therefore results in highly collimated beams of radiation. In the process, Dicke thus introduced and described the ``photon bunching'' phenomenon before it became established within the physics community \citep{HanburyBrown1956,Dicke1964}.  

\begin{figure*}
    \centering
    \begin{minipage}[t]{.4\textwidth}
    \includegraphics[trim={0 20cm 0. 0.},clip,width=\columnwidth]{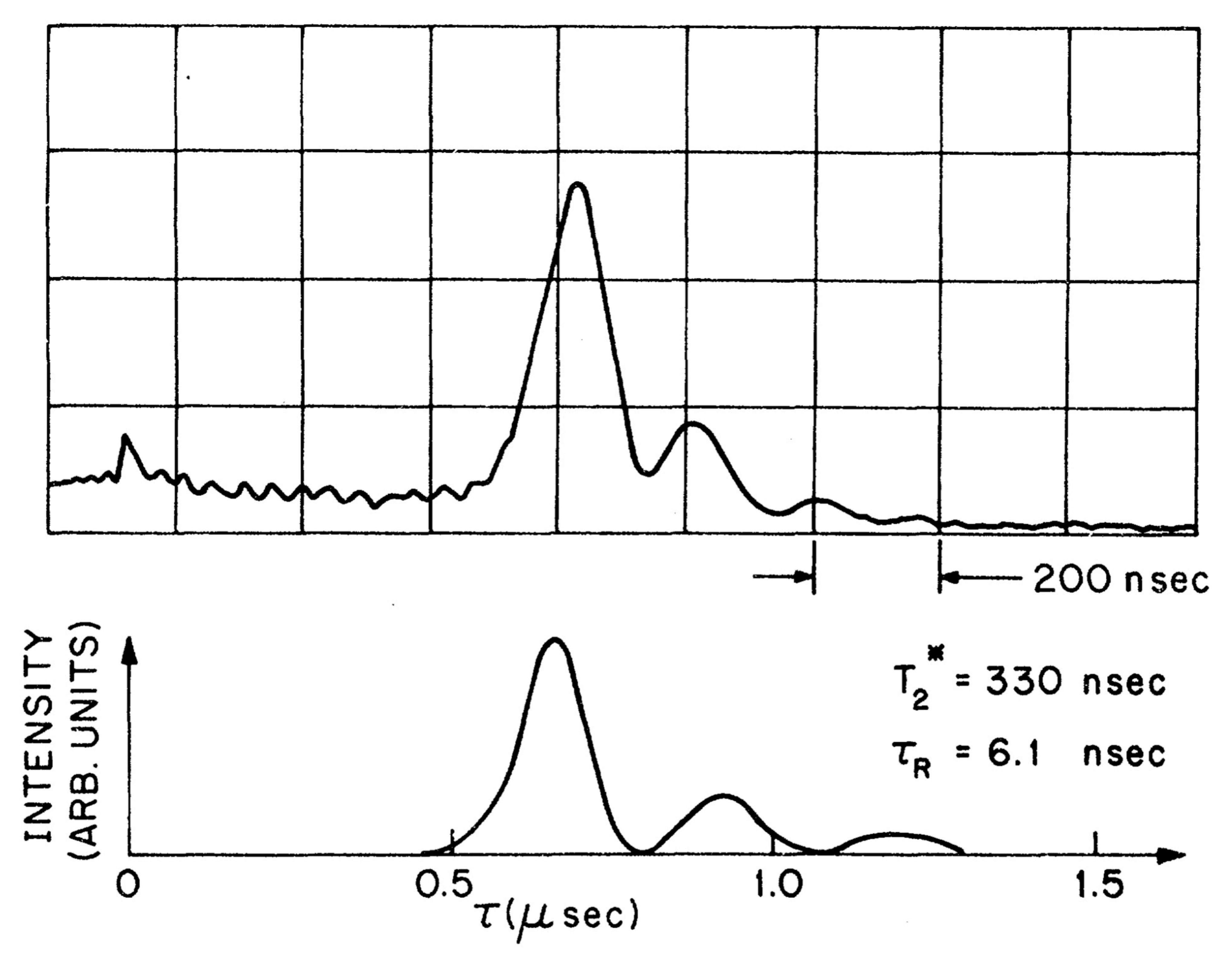}
    \end{minipage}\qquad
    \begin{minipage}[t]{.48\textwidth}
    \includegraphics[trim={0 24cm 0. 0.},clip,width=\columnwidth]{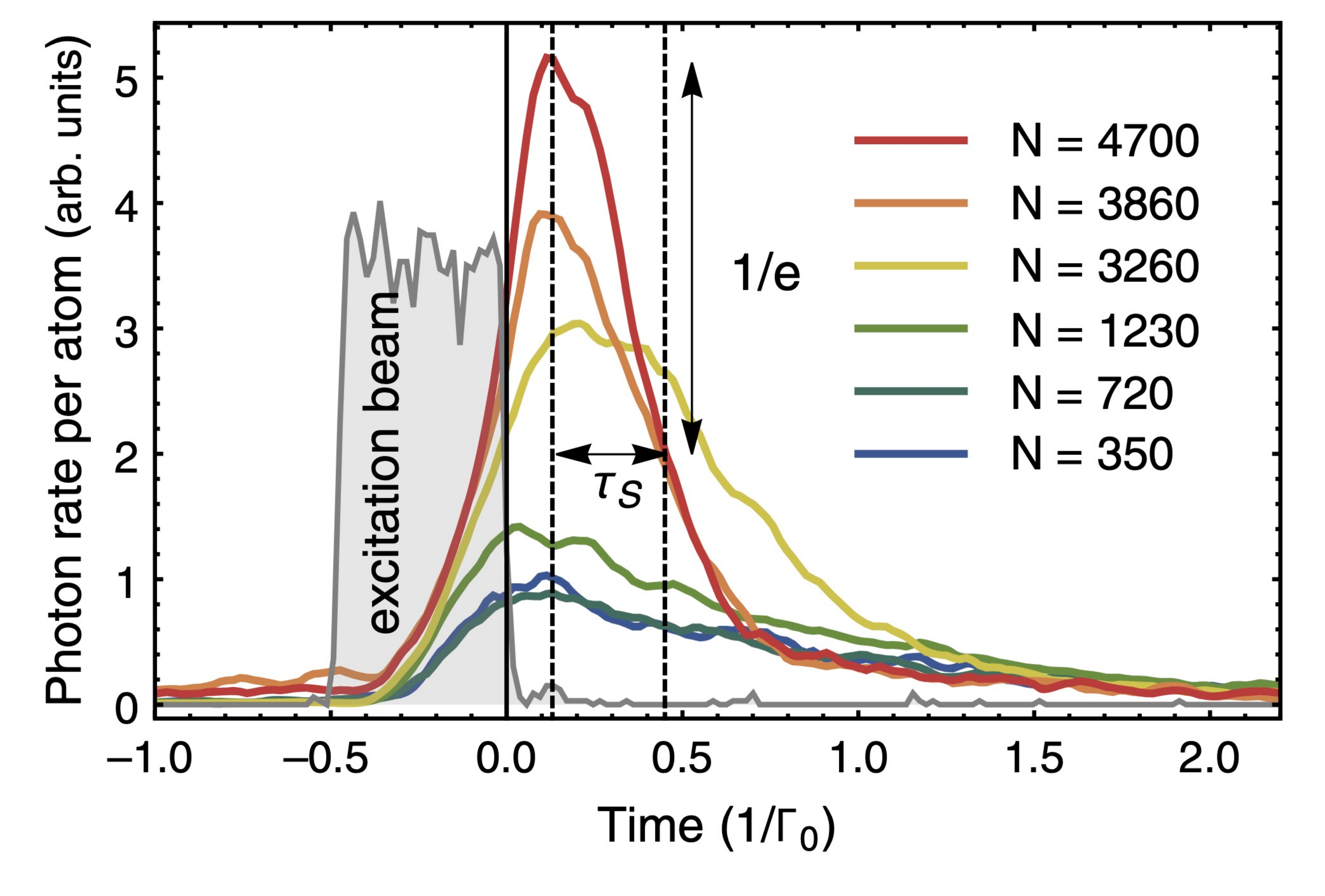}
    \end{minipage}
    \caption{\textit{Left:} First experimental realisation of superradiance by \citet{Skribanowitz1973} in a rotational transition of a vibrationally excited HF gas (top) and theoretical fit (bottom). A near complete population inversion was achieved with an intense pump pulse at $\lambda\sim 2.5\,\mu$m (the small peak at $\tau=0$ in the top graph), after which followed a superradiance burst at $84\,\mu$m. \textit{Right:} Recent laboratory results of superradiance experiments with different numbers of laser-cooled $^{87}$Rb atoms by \citet{Ferioli2021}, whose responses were probed and compared for different types of pumping conditions (the pump signal is shown in grey in the figure).}
    \label{fig:lab}
\end{figure*}

While, as previously mentioned, the theoretical introduction of superradiance preceded that of the maser in the literature, its experimental verification comparatively lagged significantly as we had to wait for almost 20 years before it could be realized in the laboratory. The left panel of Figure \ref{fig:lab} shows the first experimental verification of superradiance obtained by \citet{Skribanowitz1973} in a rotational transition of a vibrationally excited HF gas. A near complete population inversion was achieved with an intense pump pulse at $\lambda\sim 2.5\,\mu$m (the small peak at $\tau=0$ in the top graph), after which followed a superradiance burst at $84\,\mu$m. The time delay between the pump pulse signal and the intensity burst, as well as its ``ringing,'' are telling characteristics of superradiance for these experimental conditions. Superradiance has since become an area of sustained and intense research in the quantum optics community. It has been tested and used to probe fundamental physics questions for a wide range of systems under a vast array of conditions (see Chap. 2 of \citealt{Benedict1996} for an early summary). An example of a recent experiment by \citet{Ferioli2021} is shown in the right panel of Figure \ref{fig:lab}, where the superradiance responses from assemblies containing different numbers of laser-cooled $^{87}$Rb atoms were probed and compared for different types of pumping conditions.    

It is interesting to note that despite the intense amount of research accomplished in quantum optics laboratories over several decades, superradiance has remained largely unnoticed by the astronomical community until the recent work of \citet{Rajabi2016A,Rajabi2016B,Rajabi2017,Rajabi2020} (see also \citealt{Rajabi2019,Rajabi2023}).\footnote{Dicke's superradiance, as discussed here, should not be confused with ``black hole superradiance,'' which is connected to Hawking radiation; see \citealt{Brito2015} for a review of black hole superradiance and its historical connection to Dicke's superradiance.}

\section{The complementarity of the maser and superradiance regimes\label{sec:complementarity}}

\begin{figure}
    \begin{center}
    \includegraphics[trim={0 20cm 0. 0.},clip,width=0.70\textwidth]{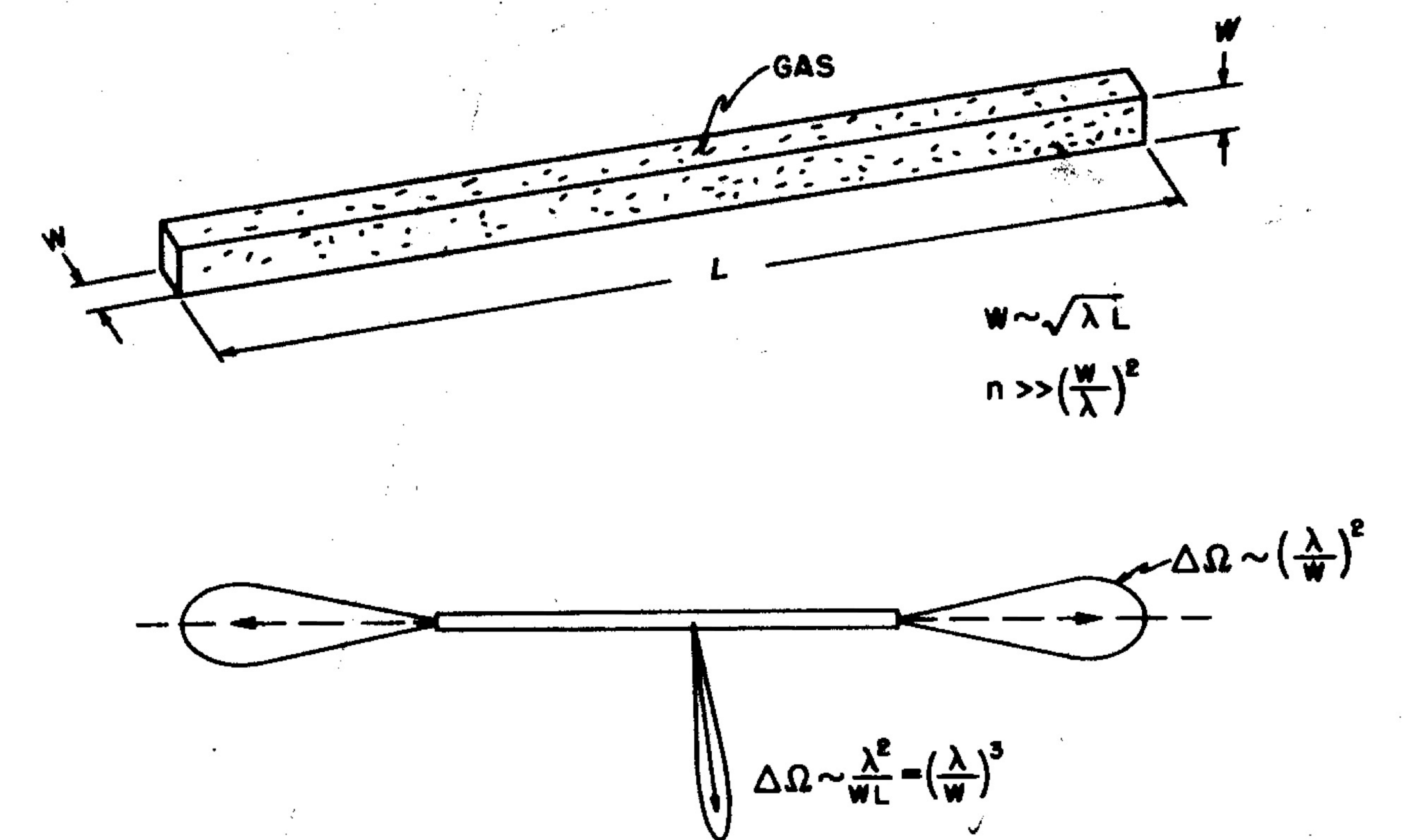}
    \caption{\textit{Top:} A cylindrical gas configuration of length $L$ containing a single molecular species at velocity coherence. The dimensions of the cylinder are consistent with a Fresnel number of unity (i.e., the width $w\sim\sqrt{\lambda L}$). \textit{Bottom:} A schematic representation of the cylinder with its diffraction patterns of solid angle $\Delta\Omega$. We will investigate the radiation emitted within one lobe at the end of the cylinder, say, on the right, with $\Delta\Omega\sim \left(\lambda/w\right)^2$. Taken from \citet{Dicke1964}.}
    \label{fig:sample-Dicke}
    \end{center}
\end{figure}

The temporal evolution of a gas, whose components (e.g., molecules) are modeled as two-level systems, interacting with an ambient electromagnetic field is described by the so-called Maxwell-Bloch equations (MBE). The application of the MBE is not limited to systems hosting a population inversion but is general and applies to a wide range of conditions where numerous fundamental physical phenomena can be theoretically studied and modeled. For the problem at hand, we will consider a gas whose (single) molecular constituents exhibit velocity coherence and occupy a linear volume configuration, e.g., a circular or rectangular cylinder as shown in Figure \ref{fig:sample-Dicke}, and focus on a single transition between two specific energy levels. 

While the MBE can in principle be numerically solved in three dimensions, it is simpler (both analytically and computationally) and adequate to limit the analysis to a one-dimensional problem \citep{Gross1982}. With this approximation the only remaining spatial variable, $z$, is that defining the symmetry axis of the cylinder while the temporal evolution is tracked as a function of the retarded time $\tau=t-z/c$. The three-dimensional reality of the problem is recovered ``by hand'' after the computations are effected by imposing a Fresnel number of unity, where the width of the cylinder is set to $w=\sqrt{\lambda L}$ with $L$ the length of the system. This condition limits the radiation intensity of the one-dimensional system to that contained to a solid angle $\Delta\Omega\sim \left(\lambda/w\right)^2$ at one end-fire of the cylinder (see the bottom schematic in Figure \ref{fig:sample-Dicke}) and restricts phase coherence to a volume of physically allowable dimensions. We will refer to such system as a ``sample.''  

Under the slowly varying envelope (SVEA) and rotating wave approximations, the one-dimensional MBE in the rest frame of the gas take the form
\begin{align}
     & \frac{\partial n^\prime}{\partial\tau} = \frac{i}{\hbar}\left(P^+E^+-E^-P^-\right)-\frac{n^\prime}{T_1}+ \Lambda_n \label{eq:dN/dt} \\
     & \frac{\partial P^+}{\partial\tau} = \frac{2id^2}{\hbar}E^-n^\prime-\frac{P^+}{T_2}+\Lambda_P \label{eq:dP/dt} \\
     & \frac{\partial E^+}{\partial z} = \frac{i\omega_0}{2\epsilon_0c} P^-\label{eq:dE/dz},
\end{align}
where $n^\prime$ is (half of) the population density difference between the upper and lower energy levels, while $P^+$ and $E^+$ are the amplitudes of the molecular polarization and the electric field, respectively defined by
\begin{align}
& \mathbf{P}^\pm\left(z ,\tau\right) = P^\pm\left(z ,\tau\right) e^{\pm i\omega_0\tau}\boldsymbol{\epsilon}_\mathrm{d}\label{eq:penvelope}\\
& \mathbf{E^\pm}\left(z ,\tau\right) = E^\pm\left(z ,\tau\right) e^{\mp i\omega_0\tau}\boldsymbol{\epsilon}_\mathrm{d}.\label{eq:Eenvelope}
\end{align}
The unit polarization vector $\boldsymbol{\epsilon}_\mathrm{d} = \mathbf{d}/d$ is associated with the molecular transition of dipole moment $d=\left| \mathbf{d}\right|$ at frequency $\omega_0$. The superscript ``$+$'' in equations (\ref{eq:dN/dt})-(\ref{eq:Eenvelope}) is for the polarization corresponding to a transition from the lower to the upper level and the positive frequency component of the electric field \citep{Gross1982,Rajabi2020,Rajabi2023}. Any non-coherent phenomena (e.g., collisions) that cause relaxation of the population density and de-phasing of the polarization are respectively accounted for through the phenomenological time-scales $T_1$ and $T_2$. The evolution of the system is initiated through internal fluctuations in $n^\prime$ and $P^+$ modeled with an initial non-zero Bloch angle
\begin{equation}
    \theta_0=\frac{2}{\sqrt{N}}, \label{eq:theta0}
\end{equation}
with $N$ the number of molecules in the sample (see \citealt{Rajabi2019} for more details). The polarization pump $\Lambda_P$ in equation (\ref{eq:dP/dt}) consists entirely of those fluctuations (i.e., for $P^\pm$). The inversion pump $\Lambda_n$ in equation (\ref{eq:dN/dt}) is for an effective pump signal that directly drives a population inversion. 

In this section we will endeavour to establish the complementarity between the maser action and superradiance, and demonstrate how they arise from the MBE. To do so, we will follow the discussion presented in Sec. 3.1 of \citet{Rajabi2020} for a system that is ``instantaneously inverted'' at $\tau=0$. That is, we set the inversion pump to 
\begin{equation}
    \Lambda_n\left(z ,\tau\right) = \left\{ 
                \begin{array}{ll}
                   0,  & \tau < 0 \\
                   \Lambda_0,  & \tau \geq 0
                \end{array}\right. \label{eq:Lambda_n_constant}
\end{equation}
with $\Lambda_0$ a constant level and fix an initial population inversion level $n^\prime_0 \equiv n^\prime\left(z,\tau=0\right)=\Lambda_0 T_1$ at all positions $z$. The inversion pump signal would then compensates for any non-coherent relaxation in the weak intensity limit (i.e., when $\partial n^\prime/\partial\tau \approx 0$), while setting a non-zero initial population inversion level is equivalent to having an instantaneous inversion at $\tau=0$.\footnote{The nature of $\Lambda_n$ will change in the next section when modeling data for G9.62+0.20E.} With this set-up, we let the system evolve and monitor $n\equiv 2n^\prime, P^+$ and $E^+$ at all positions $z$ and retarded times $\tau$.

\begin{figure*}
    \centering
    \begin{minipage}[t]{.47\textwidth}
    \includegraphics[width=\columnwidth]{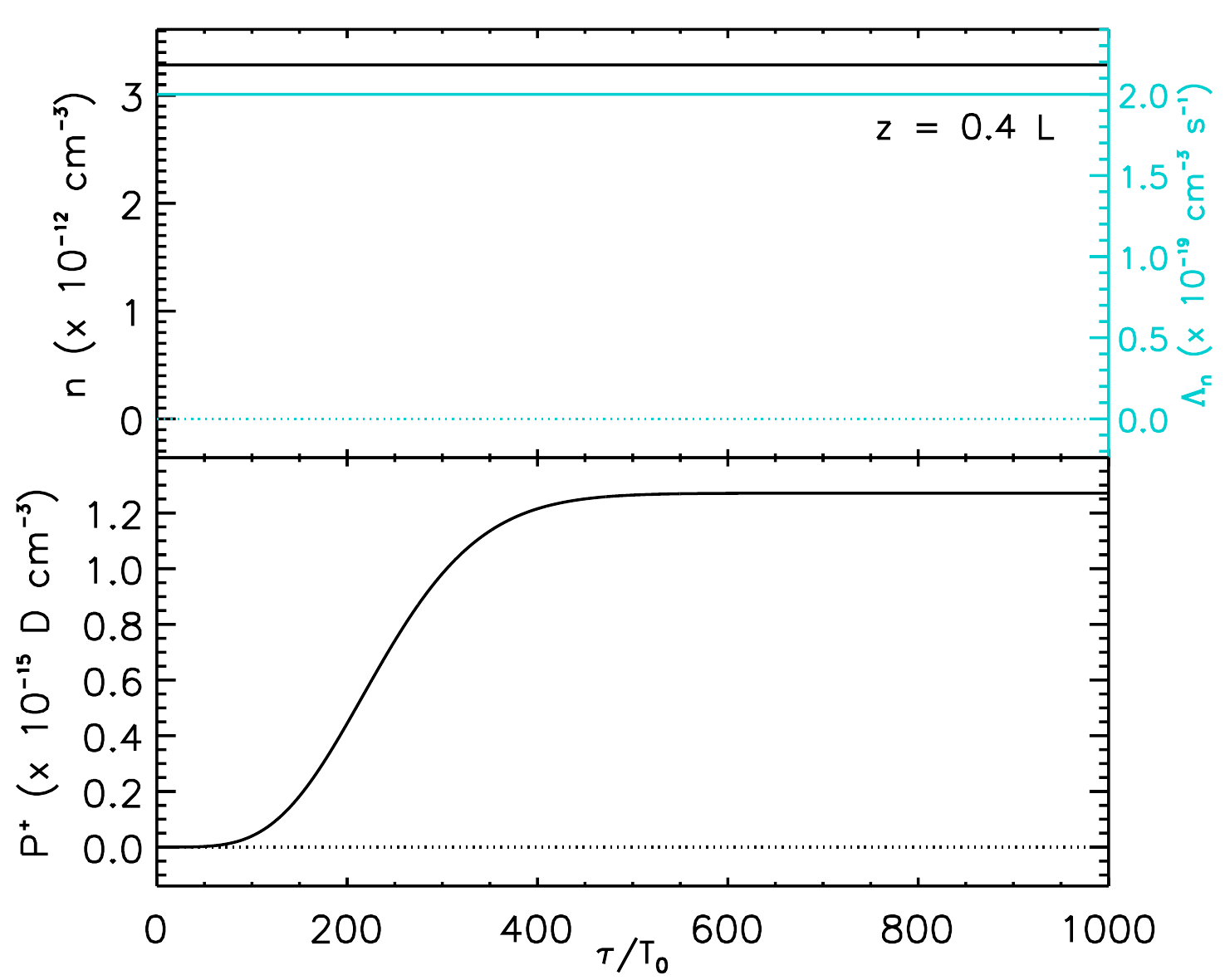}
    \end{minipage}\qquad
    \begin{minipage}[t]{.47\textwidth}
    \includegraphics[width=\columnwidth]{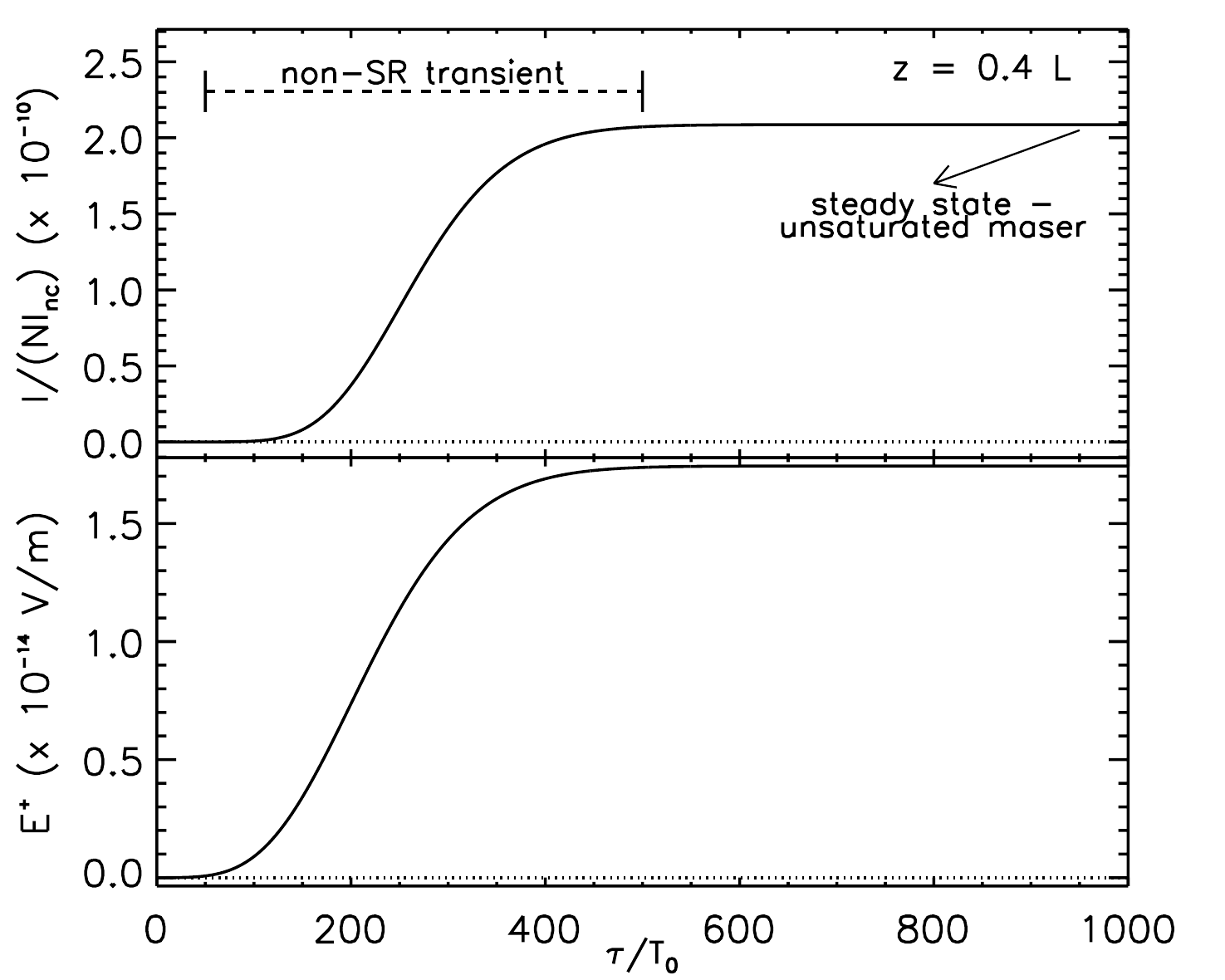}
    \end{minipage}
    \begin{minipage}[t]{.47\textwidth}
    \includegraphics[width=\columnwidth]{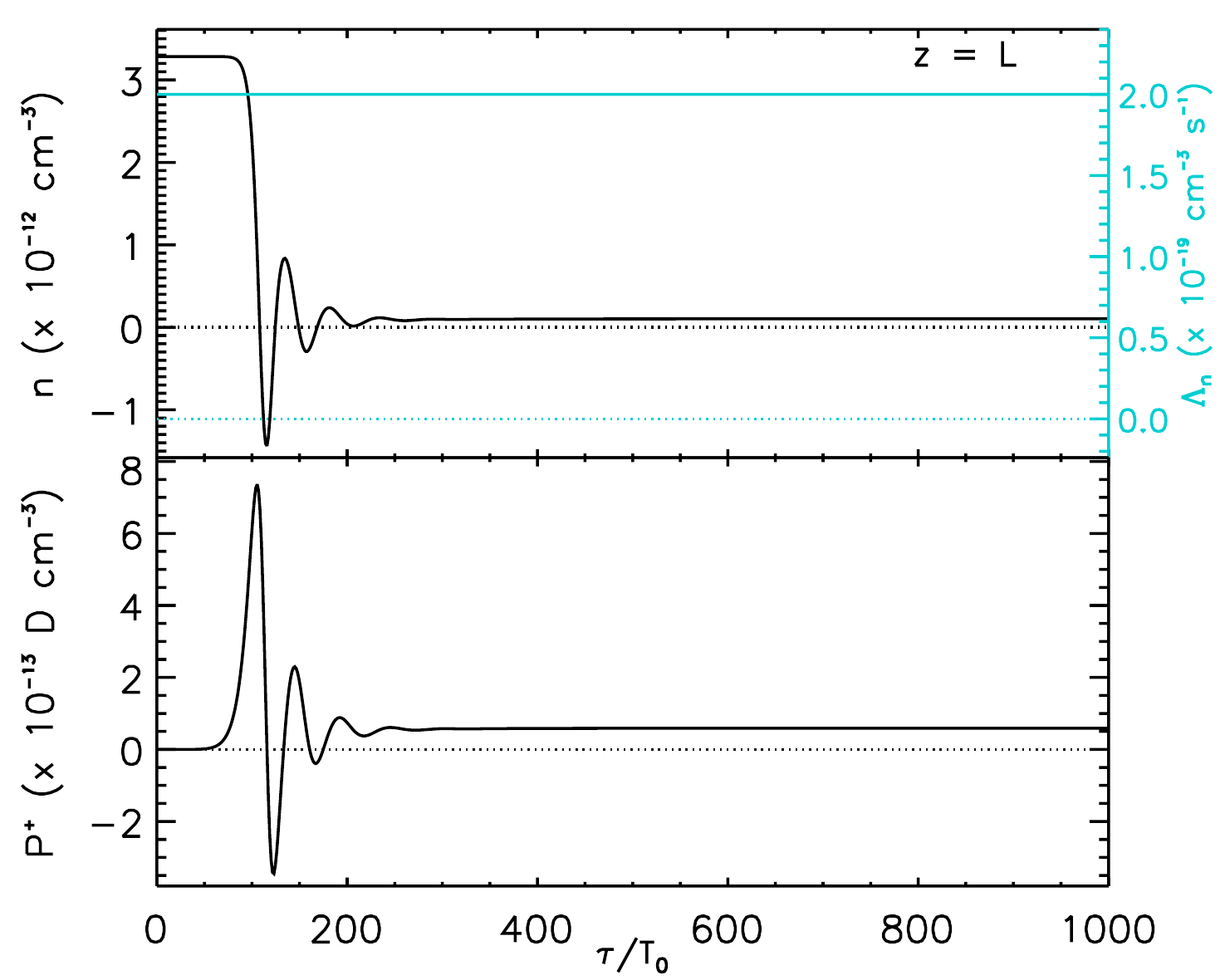}
    \end{minipage}\qquad
    \begin{minipage}[t]{.47\textwidth}
    \includegraphics[width=\columnwidth]{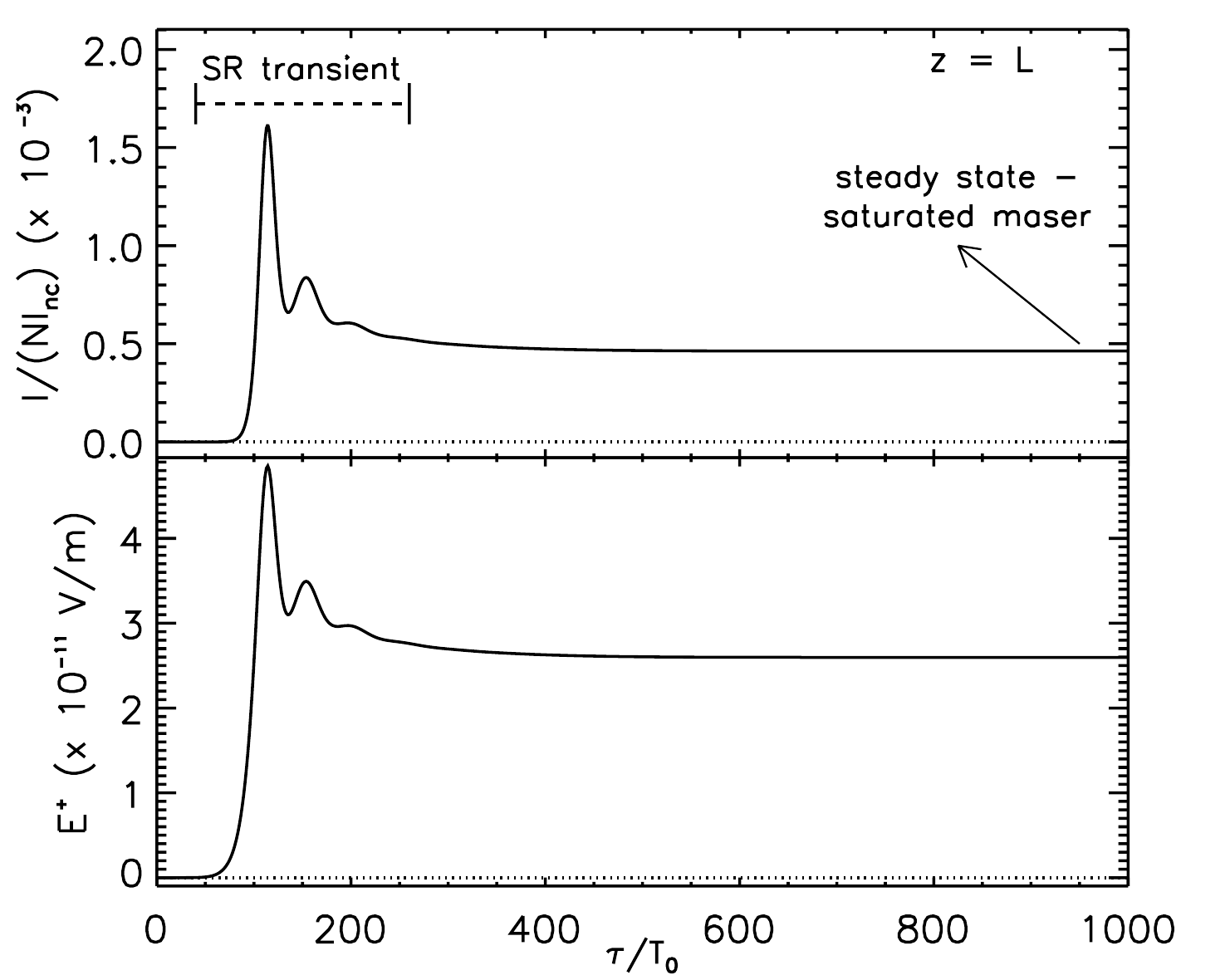}
    \end{minipage}    
    \caption{\textit{Top:} Temporal evolution of the population inversion $n\equiv 2n^\prime$ and $P^+$ (\textit{left}), and the radiation intensity $I$ and $E^+$ at $z=0.4\,L$ (\textit{right}). The intensity exhibit a smooth, non-superradiance (non-SR), transient response as the critical threshold has not been reached (see equation \ref{eq:nLcrit}). The inversion pump signal is shown in light/cyan using the vertical scale on the right and $T_0=1\times 10^5$~s. \textit{Bottom:} Same as above but at the end-fire, i.e., at $z = L$, of the sample. Here, a strong oscillatory transient regime is seen in the intensity since the inverted column density exceeds the critical threshold for superradiance. This is accompanied by an increase in coherence as assessed by the peak polarization level $P_\mathrm{max}^+\sim n_0 d$ ($d\simeq 0.7$~D), which supports a picture where the molecules in the gas behave cooperatively as a single macroscopic electric dipole. Adapted from \citet{Rajabi2020}.}
    \label{fig:saturated_step}
\end{figure*}

We show in Figure \ref{fig:saturated_step} the evolution at two positions of such a sample for a methanol gas at the 6.7~GHz spectral transition. For this example we set $L=2\times 10^{15}$~cm and $n_0\equiv 2n_0^\prime=3.3\times 10^{-12}\,\mathrm{cm}^{-3}$, which corresponds to a population inversion of approximately $0.1\,\mathrm{cm}^{-3}$ for a $1\,\mathrm{km\,s}^{-1}$ velocity distribution. These two parameters, along with the Einstein spontaneous emission coefficient for that line ($\Gamma_\mathrm{6.7\,\mathrm{GHz}}=1.56\times 10^{-9}\,\mathrm{s}^{-1}$), set the superradiance characteristic time-scale for this sample through
\begin{equation}
    T_\mathrm{R} = \frac{8\pi}{3\lambda^2 n_0L}\tau_\mathrm{sp}, \label{eq:TR}
\end{equation}
where $\tau_\mathrm{sp}=\Gamma_{\mathrm{6.7\,GHz}}^{-1}$ leads to $T_{\mathrm{R},\,\mathrm{6.7\,\mathrm{GHz}}}=4\times 10^4$~s \citep{Rajabi2020}. Finally, the time-scales for non-coherent relaxation and de-phasing were fixed to $T_1=1.64\times 10^7$~s and $T_2=1.55\times 10^6$~s, respectively. Before we study the response of the sample we note that the initial column density of the population inversion $n_0L$ appears in the denominator of $T_\mathrm{R}$, implying that an increased length for the system will lead to a faster response. This characteristic will be essential for understand the sample's behaviour and will lead to the notion of a critical threshold for the appearance of superradiance. We also note that in the figure the intensity $I=c\epsilon_0\left|E^+\right|^2/2$ is normalized to $NI_\mathrm{nc}$ with the non-coherent intensity (i.e., that expected from the sample when the molecules are radiating independently) given by 
\begin{equation}
    I_\mathrm{nc} = \frac{2}{3}\frac{\hbar\omega_0}{AT_\mathrm{R}}, \label{eq:Ync}
\end{equation}
where $A=\lambda L$ is the cross-section of the sample's end-fire (for a Fresnel number of unity; see \citealt{Rajabi2016B}).

In Figure \ref{fig:saturated_step} we see that the response of the system has an entirely different character depending on the position where it is probed. At $z=0.4\,L$ the intensity (top right) exhibits a smooth transition from $I\approx 0$ at $\tau=0$ to its steady-state value, which is very weak at $I\sim 10^{-10}NI_\mathrm{nc}$. This low intensity is due to the absence of coherence in the gas, as can also be assessed from the low level of polarization $P^+\sim10^{-15}\,\mathrm{D\,cm}^{-3}\ll n_0d$ (top left; $d\simeq 0.7$~D for this transition). For reasons that will soon become clear we will term this smooth transient regime as ``non-superradiance'' (or ``non-SR'' in the figure). The sample's response is completely different at its end-fire, i.e., at $z = L$, where a strong oscillatory transient regime is seen in the intensity, peaking at $I\simeq 1.6\times 10^{-3}NI_\mathrm{nc}$ before reaching a steady-state value of $I\simeq 0.5\times 10^{-3}NI_\mathrm{nc}$. The strong transient is also seen in $n$ and $P^+$ with the latter reaching a peak value of $P^+\sim n_0d$. This is indicative of the presence of coherence in the gas during the transient regime, which we qualify as ``SR transient.'' 

\begin{figure}
    \begin{center}
    \includegraphics[width=0.85\textwidth]{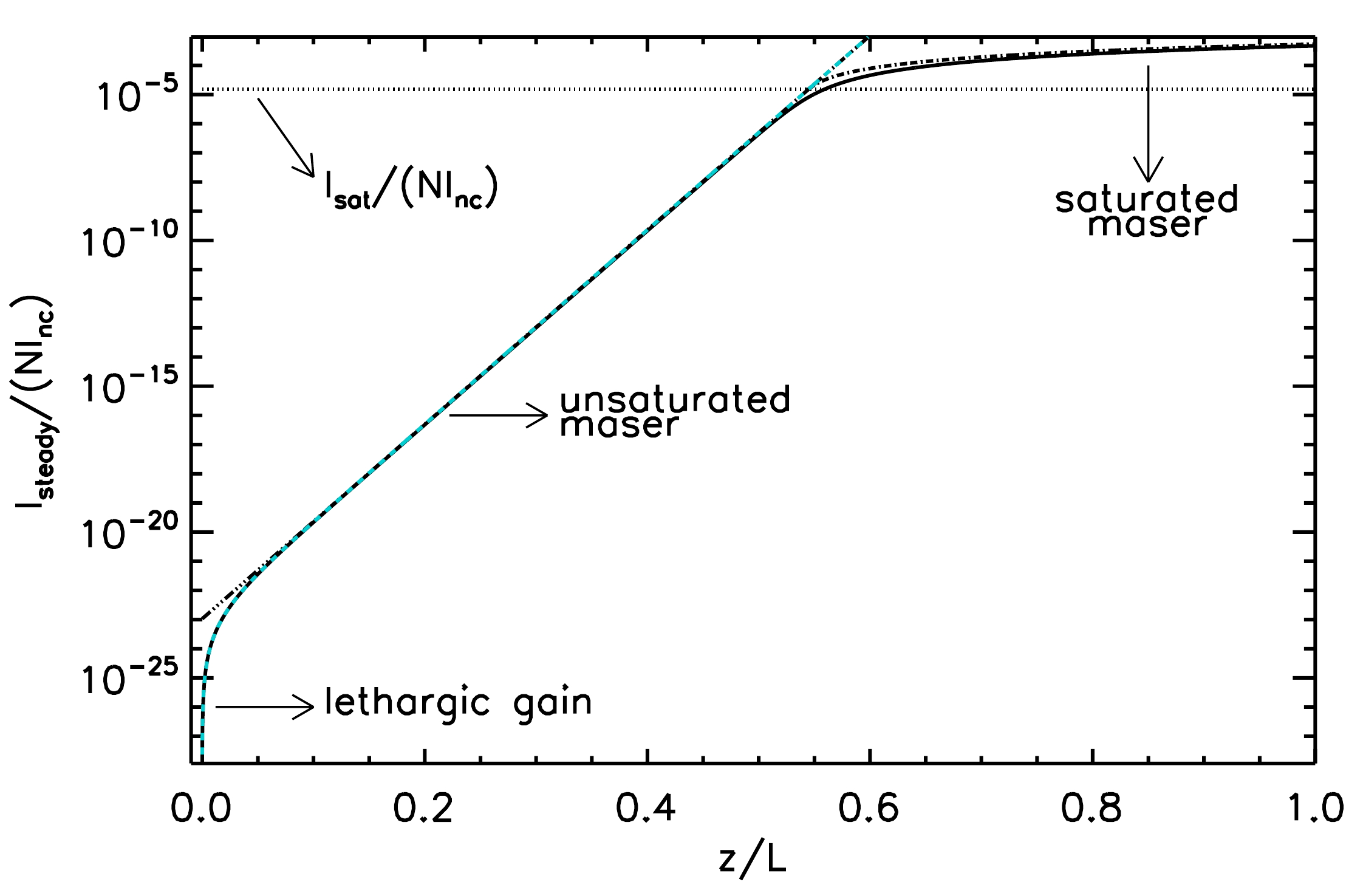}
    \caption{Steady-state (maser) intensity at the sample's end-fire ($z = L$). The unsaturated (exponential growth) and saturated (linear growth) regimes are indicated; note the logarithmic scale for the intensity. These two domains meet at the critical length $z_\mathrm{crit}/L\approx 0.54$, where the intensity neighbours the saturation intensity $I_\mathrm{sat}$ (shown with the horizontal dotted line). Taken from \citet{Rajabi2020}; see this paper for a more in-depth discussion.}
    \label{fig:maser-regimes}
    \end{center}
\end{figure}
Our usage of the ``non-SR'' and ``SR transient'' terms is based on the fact that it can be analytically shown that there exist two distinct limits or regimes discernible within the framework of the MBE for an initially inverted system such as the one considered here \citep{Feld1980,Rajabi2020}. In the limit when the population density and the polarization are changing on a time-scale much shorter than $T_1$ and $T_2$, i.e.,  
\begin{equation}
\frac{\partial n^\prime}{\partial\tau} \gg \frac{n^\prime}{T_1} \quad\mathrm{and} \quad \frac{\partial P^+}{\partial\tau} \gg \frac{P^+}{T_2}, \label{eq:transientlimit}
\end{equation}
the sample enters a rapid transient regime that is characterized by the presence of coherence in the gas and strong intensities scaling with $N^2$ (see below). This is the domain of Dicke's superradiance. The bottom graphs in Figure \ref{fig:saturated_step} form indeed a good example of an ``SR transient.'' The slower, non-coherent transient regime observed in the top graphs in Figure \ref{fig:saturated_step} are thus ``non-SR transient.''   

On the other hand, at the opposite limit when the population density and the polarization are varying on a time-scale much longer than $T_1$ and $T_2$, i.e.,  
\begin{equation}
\frac{\partial n^\prime}{\partial\tau} \ll \frac{n^\prime}{T_1} \quad\mathrm{and} \quad \frac{\partial P^+}{\partial\tau} \ll \frac{P^+}{T_2}, \label{eq:steadystatelimit}
\end{equation}
it can be shown that the MBE simplify to the maser equation (see Sec. 2.1 of \citealt{Rajabi2020} for a derivation). This is the so-called ``steady-state limit'' where the maser action (and stimulated emission) is at play. This steady-state regime is identified Figure \ref{fig:saturated_step} in the intensity graphs on the right. More precisely, at $z=0.4\,L$ (top) the low intensity suggests the existence of a (steady-state) unsaturated maser while at $z=L$ (bottom) we are in the presence of a saturated maser.

The fact that the slow limit characterized by equation (\ref{eq:steadystatelimit}) is associated with the maser action can be readily verified by plotting the (normalized) steady-state intensity $I_\mathrm{steady}/\left(NI_\mathrm{nc}\right)\equiv I\left(\tau=1000T_0\right)/\left(NI_\mathrm{nc}\right)$ for all positions $z$, as shown in Figure \ref{fig:maser-regimes}. We can clearly verify this association by the presence of the exponential growth unsaturated maser and linear gain saturated maser regimes, which also justify our earlier attributions for $z=0.4\,L$ and $z=L$.   

\begin{figure}
    \begin{center}
    \includegraphics[width=0.85\textwidth]{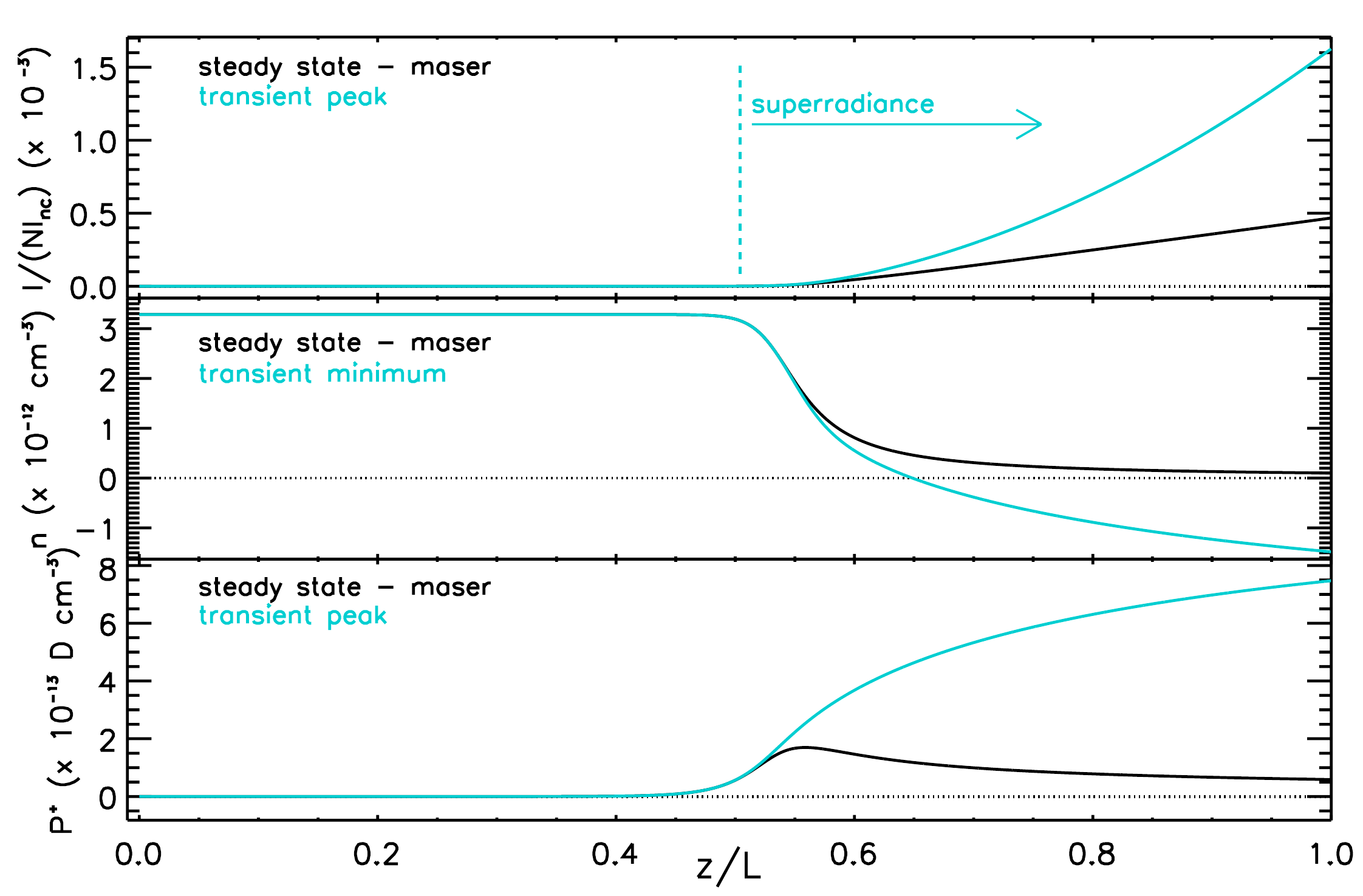}
    \caption{The superradiance/transient (light/cyan curves) and maser/steady-state (dark/black curves) intensities (top), population densities (middle) and polarization amplitudes (bottom) at all positions $z/L$ along the sample. The superradiance regime begins at $z/L\approx 0.5$ where the critical inverted column density threshold is reached; this domain also corresponds to that of the saturated maser regime. The superradiance peak intensity scales as $z^2$ while the saturated maser's is linear with position ($\propto z$). The superradiance population density minimum level reaches more negative values with increasing $z$ while the maser inversion asymptotically tends to zero. The level of coherence (i.e., the polarization amplitude) increases with distance for superradiance. Adapted from \citet{Rajabi2020}.}
    \label{fig:sample-track}
    \end{center}
\end{figure}
We can also verify the association between Dicke's superradiance and the fast transient limit by tracking the peak intensity, the minimum population density and the peak polarization as a function of the position $z$ along the sample. This is shown in Figure \ref{fig:sample-track}, where these quantities are respectively plotted from top to bottom in light/cyan along with the corresponding steady-state (i.e., maser) values in black. In the top graph for the intensities we can clearly see the linear gain for the saturated maser intensity $I_\mathrm{steady}\propto z\propto N$ and the quadratic behaviour for the transient superradiance regime $I_\mathrm{SR}\propto z^2\propto N^2$. This functionality, as well as those for the population density and the polarization, are the hallmark of Dicke's superradiance.

One more important piece of information can be gathered from Figures \ref{fig:maser-regimes} and \ref{fig:sample-track}. That is, it can be seen through their comparison that the transition between the unsaturated and saturated maser regimes (where $I_\mathrm{steady}=I_\mathrm{sat}$ in Figure \ref{fig:maser-regimes}, with $I_\mathrm{sat}$ the saturation intensity) and the appearance of superradiance (from the top graph of Figure \ref{fig:sample-track}) happen at the same position $z\equiv z_\mathrm{crit}$. It can be shown that this position is associated with the reaching of a critical threshold for the column density 
\begin{equation}
    n_0z_\mathrm{crit} = \frac{4\pi}{3\lambda^2}\frac{\tau_\mathrm{sp}}{T_2}\ln{\left(\frac{T_2}{T_1\theta_0^2}\right)},\label{eq:nLcrit}
\end{equation}
which sufficiently reduces the superradiance characteristic time-scale $T_\mathrm{R}$ to reach the fast limit of equation (\ref{eq:transientlimit}) \citep{Rajabi2020}. In other words, the appearances of superradiance in the transient response and a saturated maser in the steady-state regime are closely linked and will take place whenever $L\geq z_\mathrm{crit}$. Systems of shorter lengths will not host the superradiance phenomenon and will be restricted to the unsaturated maser regime in the steady state. We can also combine equations (\ref{eq:TR}) and (\ref{eq:nLcrit}) to reformulate this critical threshold as 
\begin{equation}
    T_\mathrm{R,crit} = \frac{2T_2}{\ln\left[T_2/\left(T_1\theta_0^2\right)\right]}. \label{eq:TRcrit}
\end{equation}
For our system we find $T_\mathrm{R}<T_\mathrm{R,crit}\approx 0.05\,T_2\simeq7\times 10^4$~s, which satisfies the requirement for superradiance.

\section{Modeling of G9.62+0.20E\label{sec:G9.62}}

\begin{figure*}
    \centering
    \begin{minipage}[t]{.47\textwidth}
    \includegraphics[width=\columnwidth]{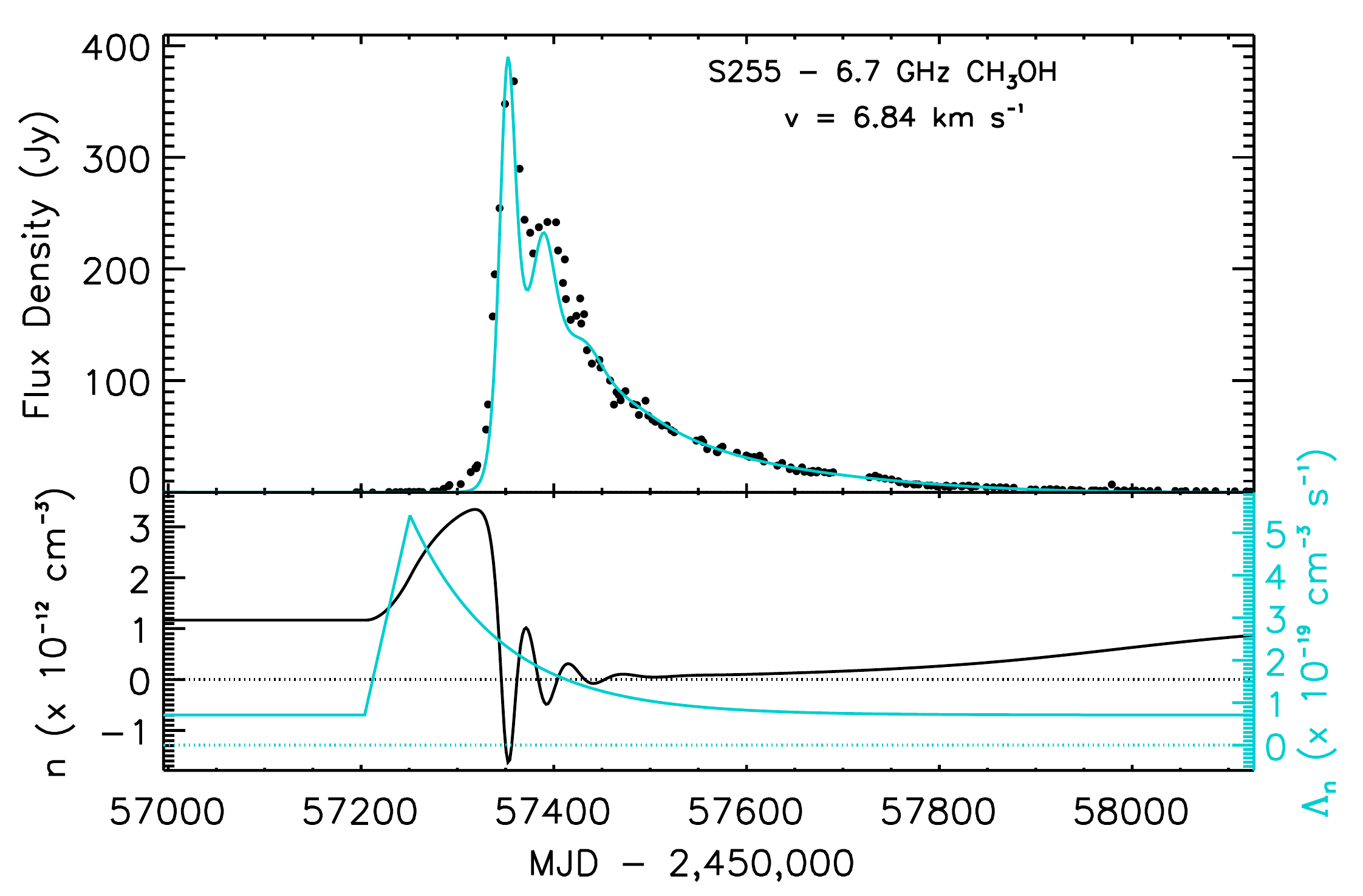}
    \end{minipage}\qquad
    \begin{minipage}[t]{.47\textwidth}
    \includegraphics[width=\columnwidth]{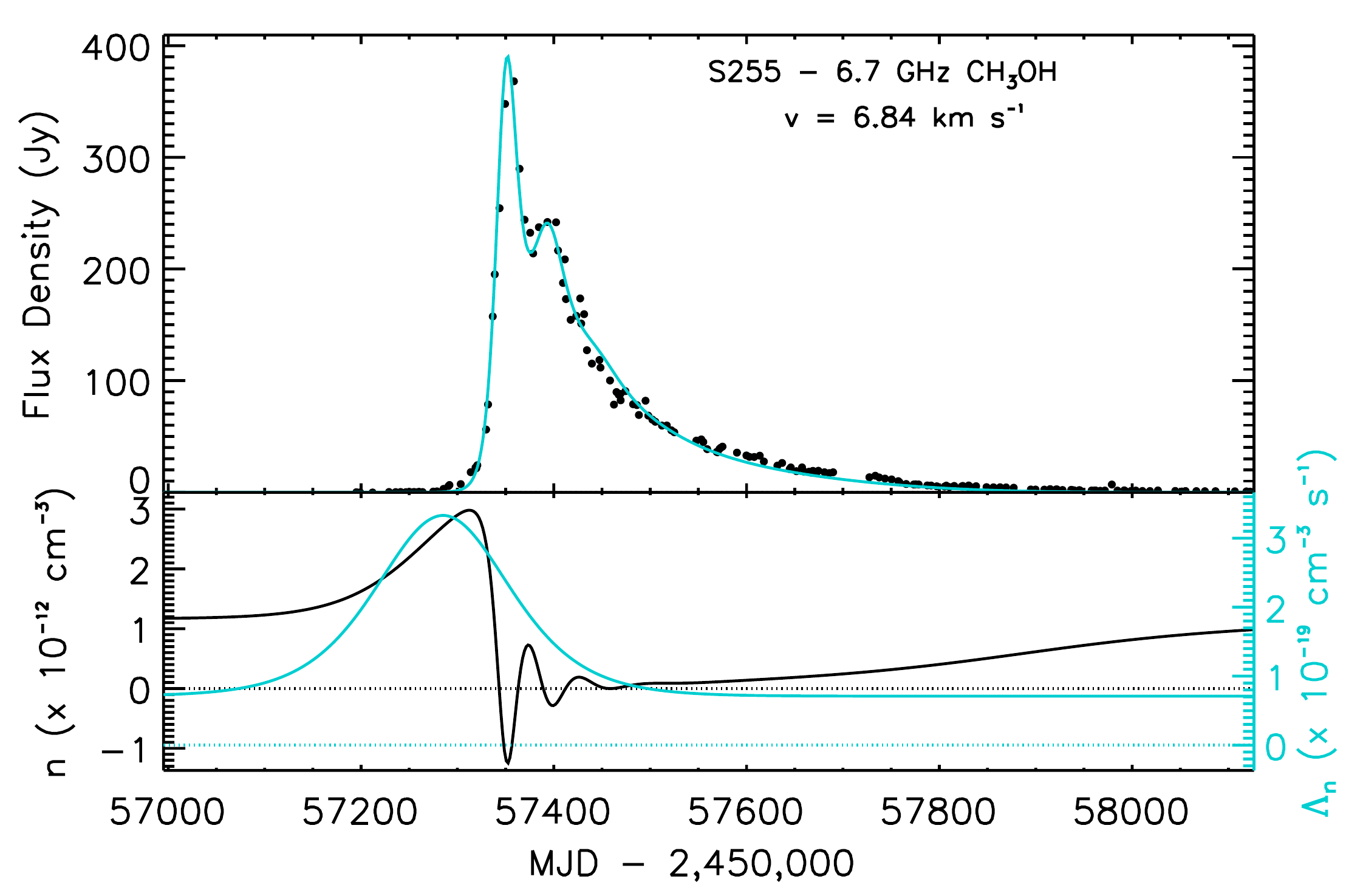}
    \end{minipage}
    \caption{Superradiance model for the S255IR-NIRS3 6.7~GHz methanol flare at $v_{\mathrm{lsr}} = 6.84~\mathrm{km~s}^{-1}$ \citep{Szymczak2018b,Rajabi2019,Rajabi2020}. The top panels show the superradiance fit (light/cyan) to the data (dots), while the bottom panels are for the pump (light/cyan, vertical axis on the right) and the population density (black, vertical axis on the left). \textit{Left:} The pump pulse is modeled with a slow rising exponential function to approximately mimic the IR observations presented in \citet{Uchiyama2020}. \textit{Right:} A smoother and symmetric function is used for the pump signal. The shape of fast transient superradiance response is thus seen to be largely independent of that of the pump signal. Taken from \citet{Rajabi2023}.}
    \label{fig:S255}
\end{figure*}

One further characteristic of the superradiance transient response not discussed in Sec. \ref{sec:complementarity} is that the shape of the intensity curve as a function of time is largely independent of that of the excitation (i.e., the pump signal or the seed electric field). An example is shown in Figure \ref{fig:S255} where a strong 6.7~GHz methanol flare in S255IR-NIRS3 is modeled using two different inversion pump signals \citep{Szymczak2018b,Rajabi2019,Rajabi2020}. Other parameters for the model (i.e., $L, T_1$ and $T_2$) are similar to those used in Sec. \ref{sec:complementarity}. We thus see that the shape of the fast transient superradiance response is minimally affected by the detailed nature of the pump signal. This behaviour will be advantageous in this section as we endeavour to model existing data from the G9.62+0.20E star-forming region. That is, while we do not know the nature of the excitation responsible for the periodic flaring observed in this source, the fact that the the shape of the measured light curves are the results of the natural (or characteristic) transient responses of the corresponding systems will allow us to perform meaningful comparisons between spectral transitions from different molecular species.

G9.62+0.20E is a high mass star-forming region located 5.2~kpc away \citep{Sanna2009}. Methanol 6.7~GHz monitoring by \citet{Goedhart2003} (and subsequent studies) has revealed this source to be flaring with a main period of approximately 243~d (a second period of 52~d was recently discovered by \citealt{MacLeod2022}). This characteristic has since led to a significant number of studies, observational and theoretical, which resulted in increased monitoring with other molecular species/transitions and models aimed at explaining the source of the periodicity as well as the physical mechanisms behind the flaring activity (see \citealt{Rajabi2023} for a recent review). The availability of multi-transition monitoring data for G9.62+0.20E is particularly interesting to us, as it will allow us to probe the transient (i.e., flaring) regimes for the different molecular species and transitions, as well as apply and test the model based on the MBE discussed in Sec. \ref{sec:complementarity}. Here, we will present some of the results from \citet{Rajabi2023}. 

The data sets we used were published elsewhere in the literature prior to the work of \citet{Rajabi2023}. That is, the OH~1665~MHz and 1667~MHz data were obtained with KAT-7 \citep{Foley2016} and initially published in \citet{Goedhart2019}, as were the methanol 12.2~GHz observations from the HartRAO 26-m telescope. The methanol 6.7~GHz observations are those from \citet{MacLeod2022}, which resulted from the corresponding monitoring of G9.62+0.20E performed with the Hitachi 32~m telescope of the Ibaraki station from the NAOJ Mizusawa VLBI Observatory \citep{Yonekura2016}. The reader is referred to the cited references for more details concerning these data and to \citet{Rajabi2023} for a more comprehensive discussion of our analysis.

\begin{figure}
    \centering
    \includegraphics[width=0.65\textwidth]{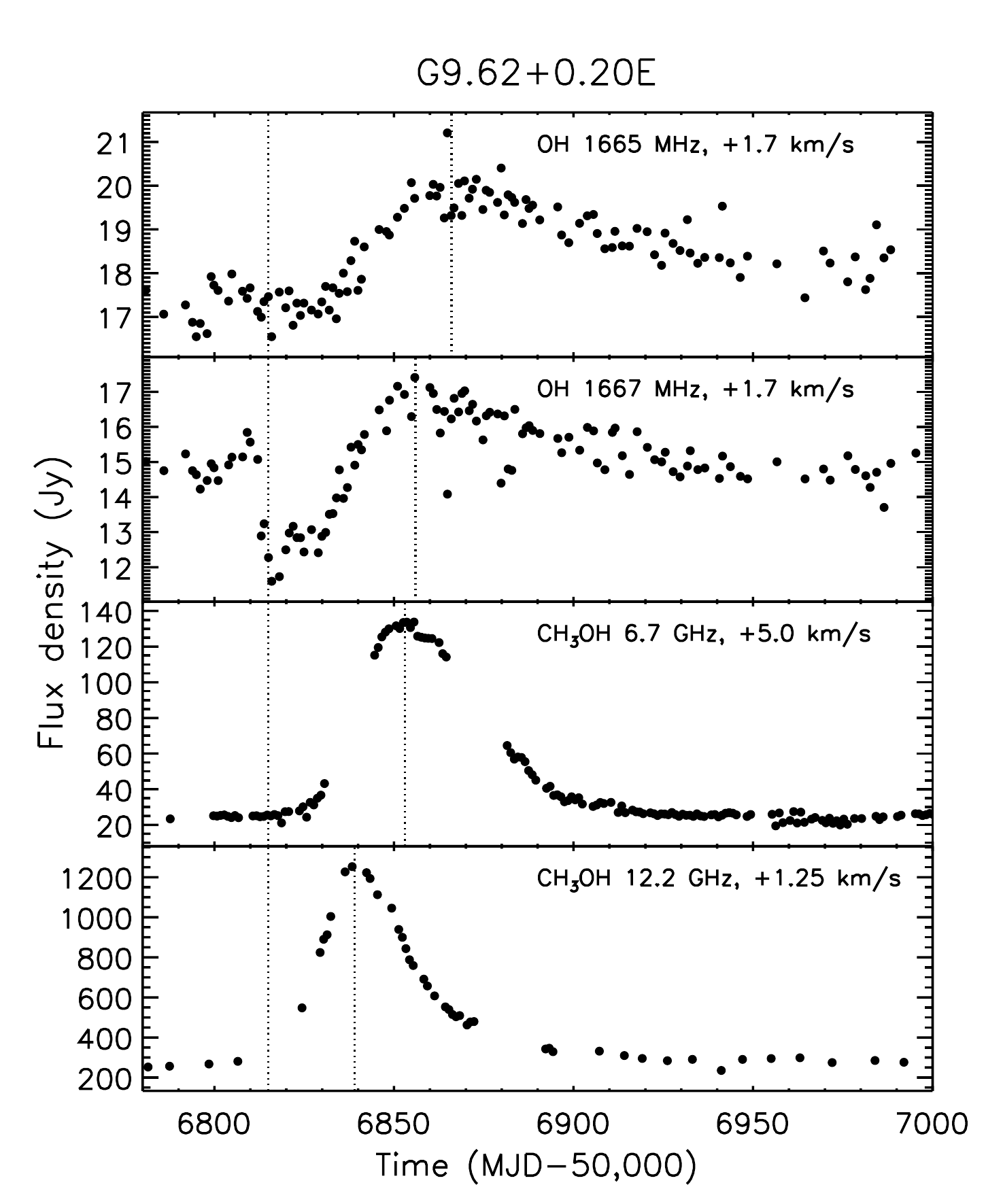}
    \caption{Comparison of profiles between OH and methanol flares in G9.62+0.20E. The vertical line at MJD 56815 denotes the start of the methanol 12.2~GHz flare, as defined by \citet{Goedhart2019}. The other vertical lines indicate the occurrence of peak intensities for each transition. The methanol 6.7~GHz light curve was delayed by 15 days to align its start with that of the others. Taken from \citet{Rajabi2023}.}
    \label{fig:g9.62-all}
\end{figure}
One interesting feature resulting from the multi-transition monitoring data of G9.62+0.20E uncovered by \citet{Goedhart2019} is the existence of different duration for the flares from different molecular species and transitions. An example is shown in Figure \ref{fig:g9.62-all} where one flare from OH 1665 and 1667~MHz and methanol 6.7 and 12.2~GHz each is shown. We can clearly see from the position of the peaks that the flare duration decreases from top to bottom in the figure (i.e., from OH 1665~MHz $\rightarrow$ OH 1667~MHz $\rightarrow$ methanol 6.7~GHz $\rightarrow$ methanol 12.2~GHz). At first sight, this behaviour may seem difficult to explain as the sources for these features are likely to be subjected to a common excitation (e.g., these flares are all recurrent at the same period of 243.3~d). However, this is the kind of results we should expect within the framework of the MBE, in view of the existence of the superradiance transient regime.

More precisely, using equation (\ref{eq:TR}) we can estimate the relative superradiance time-scales for the different spectral lines pertaining to Figure \ref{fig:g9.62-all} to be
\begin{align}
    \frac{T_\mathrm{R,1665\,MHz}}{T_\mathrm{R,6.7\,GHz}} & \simeq 1.4\frac{\left(nL\right)_\mathrm{6.7\,GHz}}{\left(nL\right)_\mathrm{1665\,MHz}}\label{eq:TR-ratio-1665_6.7}\\    
    \frac{T_\mathrm{R,1667\,MHz}}{T_\mathrm{R,6.7\,GHz}} & \simeq 1.3\frac{\left(nL\right)_\mathrm{6.7\,GHz}}{\left(nL\right)_\mathrm{1667\,MHz}}\label{eq:TR-ratio-1667_6.7}\\
    \frac{T_\mathrm{R,12.2\,GHz}}{T_\mathrm{R,6.7\,GHz}} & \simeq 0.6\frac{\left(nL\right)_\mathrm{6.7\,GHz}}{\left(nL\right)_\mathrm{12.2\,GHz}},\label{eq:TR-ratio-12.2_6.7}
\end{align}
which, interestingly, scale in the same manner as the observations for equal inversion column densities. Of course, there is no guarantee that all sources will share the same value for the inversion column density but, still, this pattern can certainly serve as a motivation for verifying if our MBE/superradiance approach can help elucidate these observational results.

\begin{figure*}
    \centering
    \begin{minipage}[t]{.47\textwidth}
    \includegraphics[width=\columnwidth]{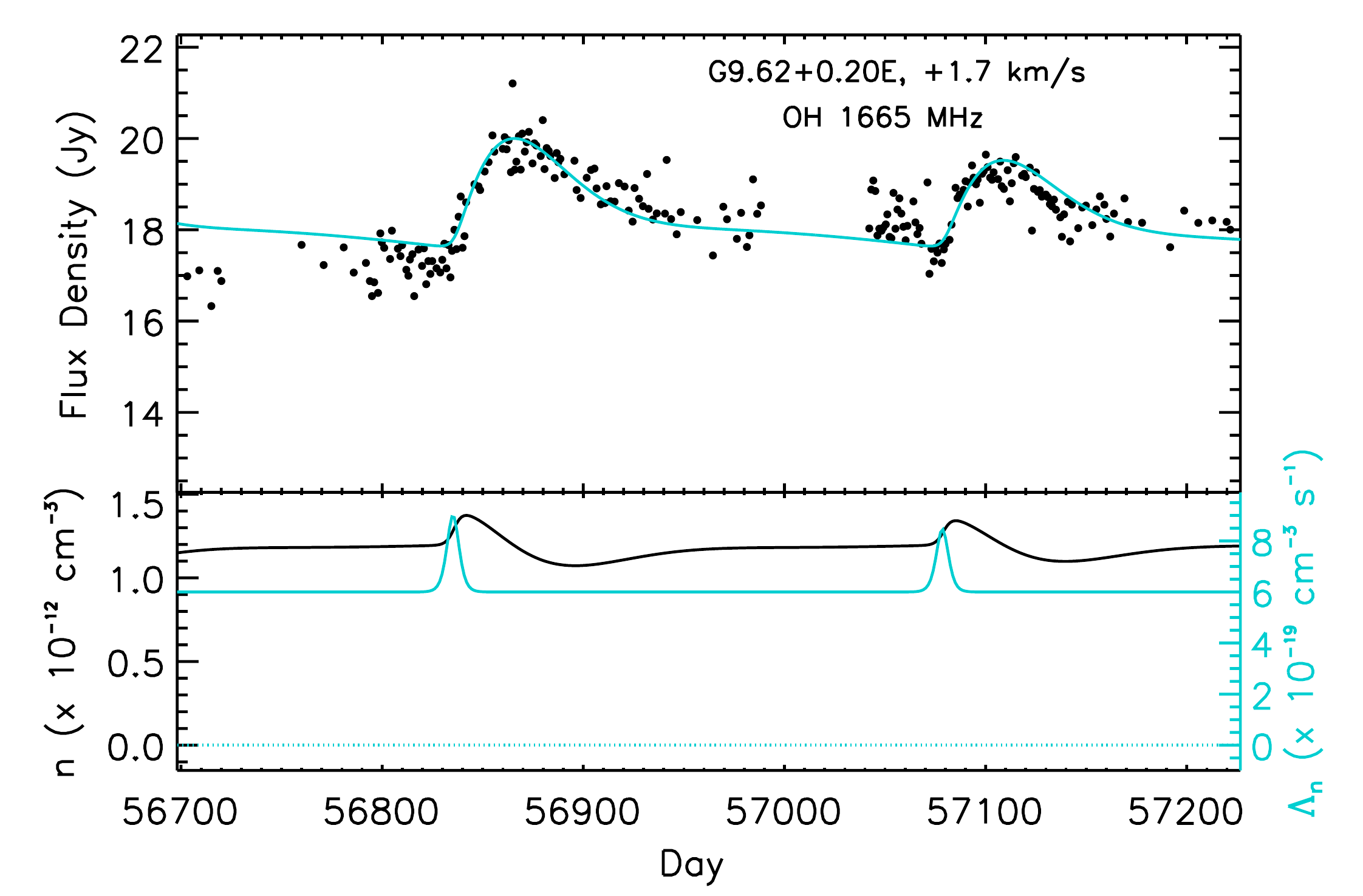}
    \end{minipage}\qquad
    \begin{minipage}[t]{.47\textwidth}
    \includegraphics[width=\columnwidth]{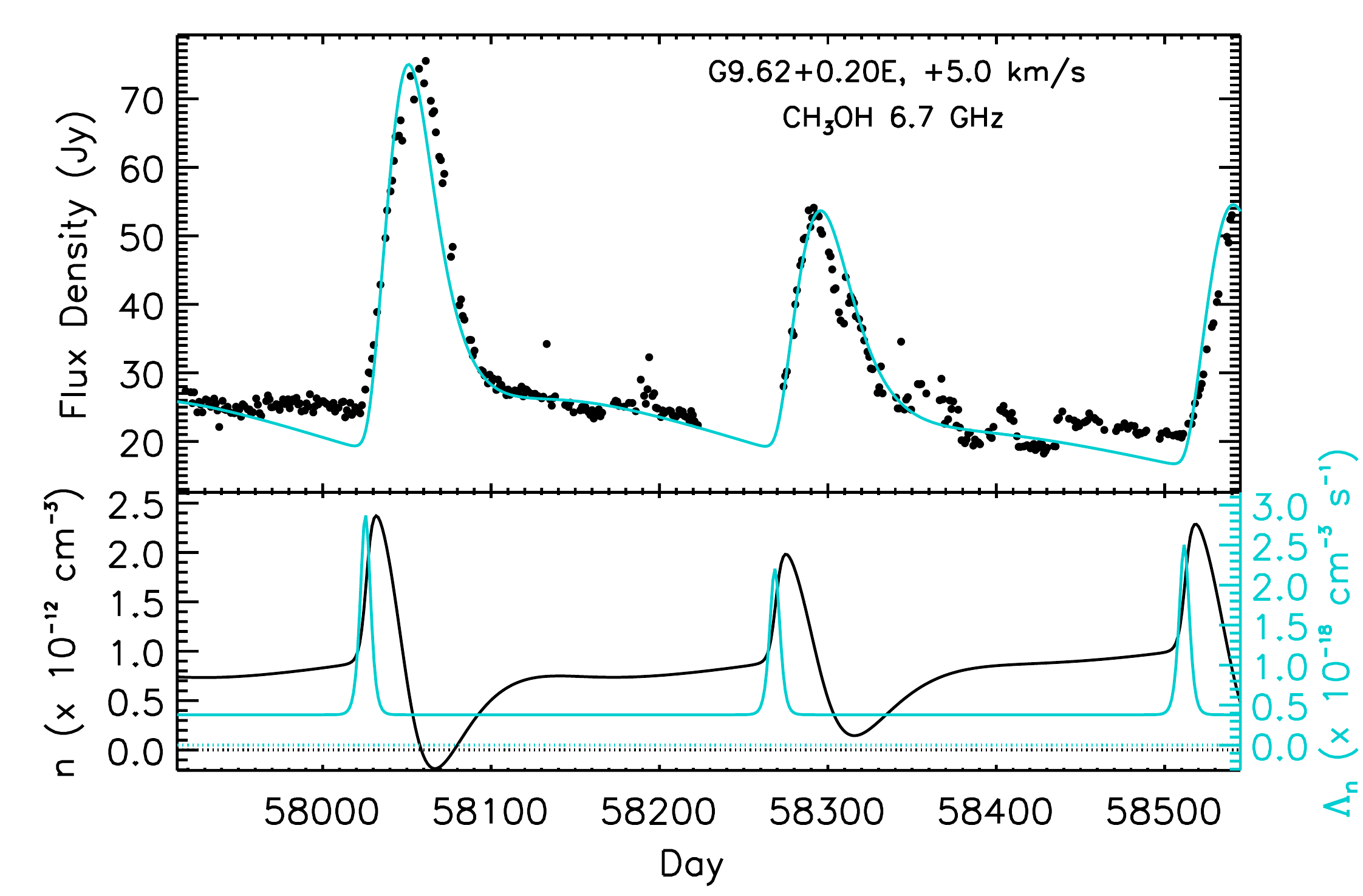}
    \end{minipage}
    \begin{minipage}[t]{.47\textwidth}
    \includegraphics[width=\columnwidth]{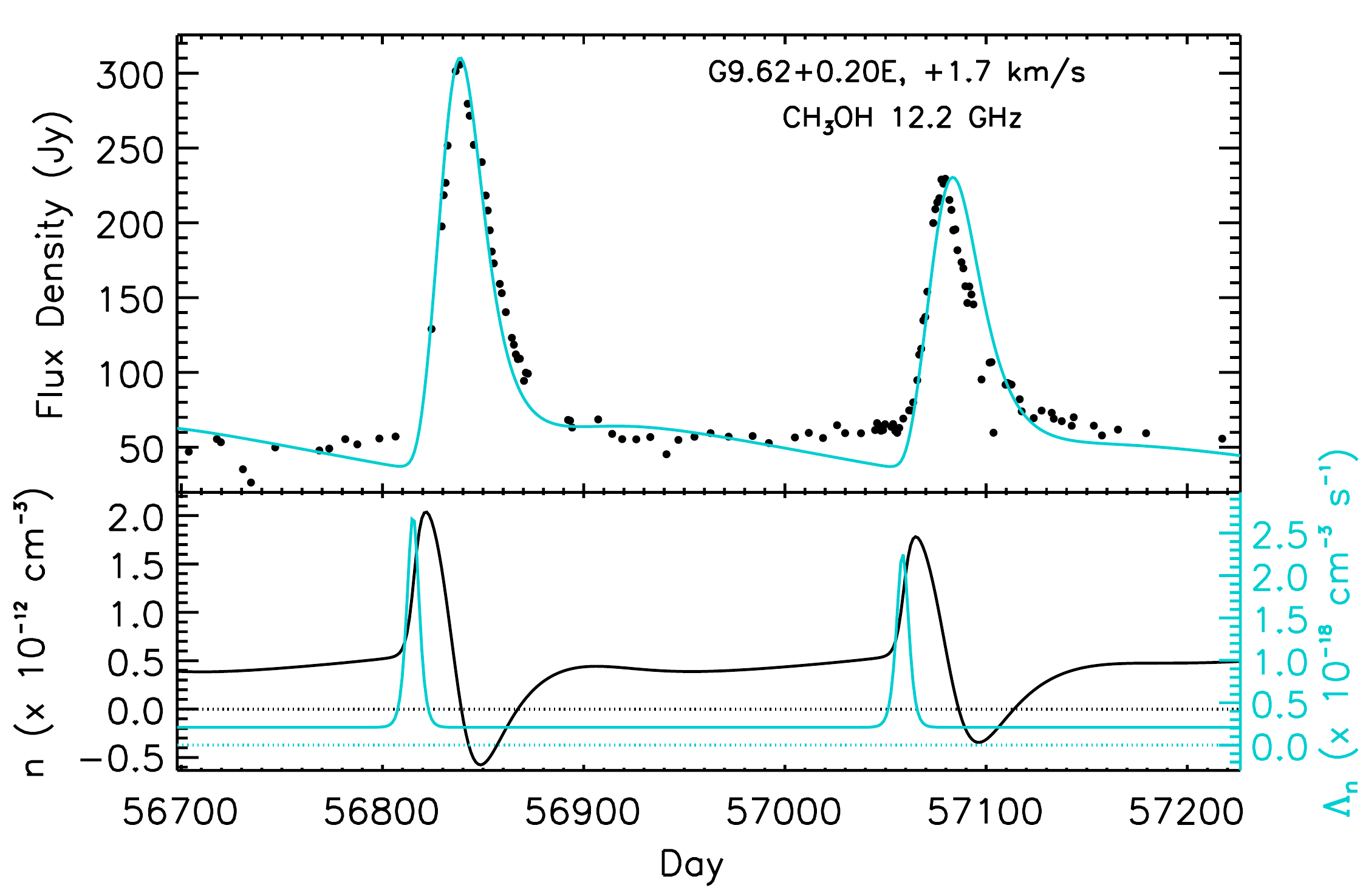}
    \end{minipage}\qquad
    \begin{minipage}[t]{.47\textwidth}
    \includegraphics[width=\columnwidth]{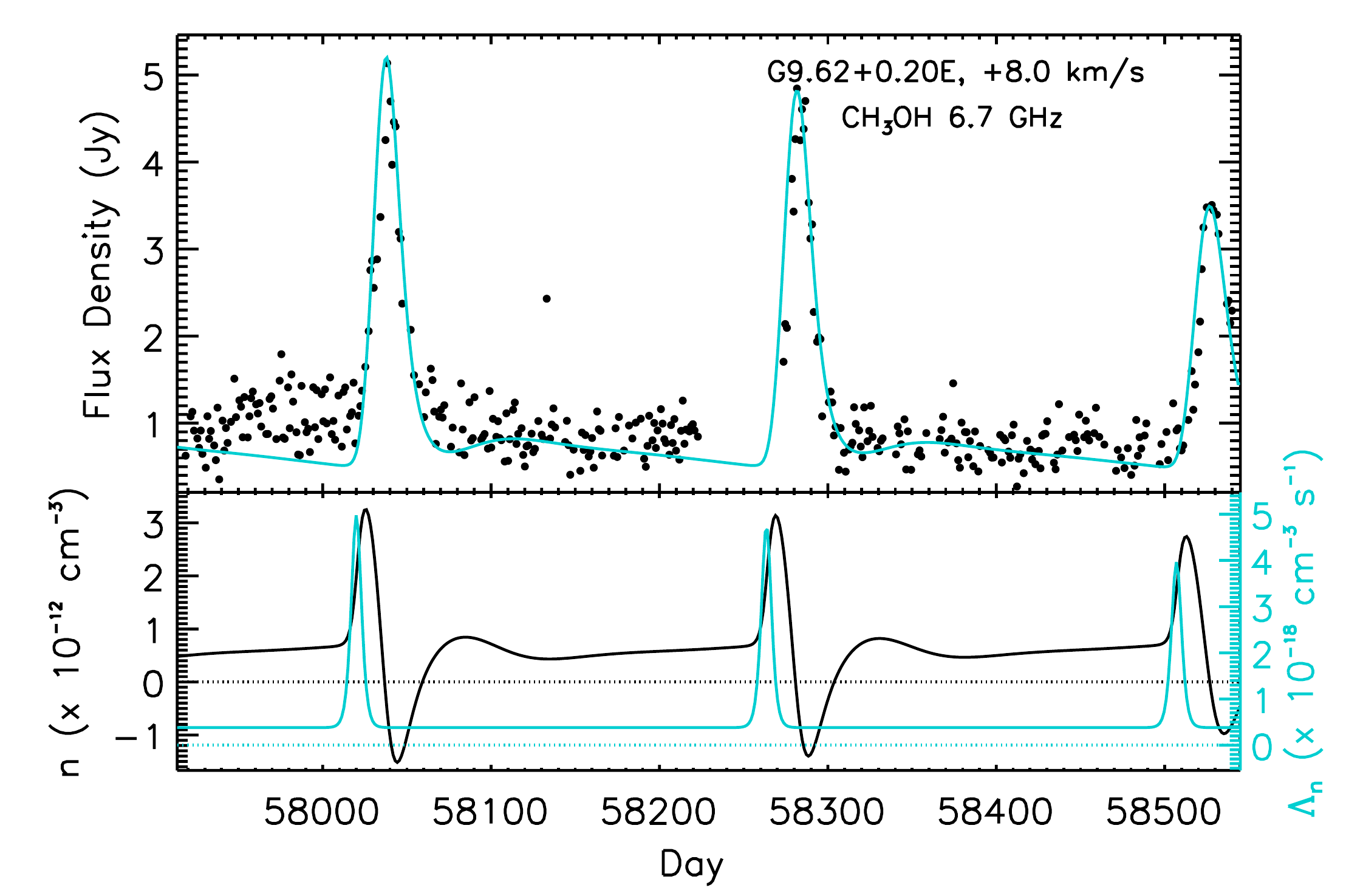}
    \end{minipage}    
    \caption{Model fits using the MBE for flaring features in OH 1665~MHz ($v_{\mathrm{lsr}} = +1.7\,\mathrm{km\,s}^{-1}$; top left), methanol 12.2~GHz ($v_{\mathrm{lsr}} = +1.7\,\mathrm{km\,s}^{-1}$; bottom left) and methanol 6.7~GHz ($v_{\mathrm{lsr}} = +5.0\,\mathrm{km\,s}^{-1}$ and $+8.0\,\mathrm{km\,s}^{-1}$; top and bottom right, respectively). Each graph shows the model fit (light/cyan; top) to the data (black dots), the population density (black curve; bottom, left vertical scale) and the inversion pump signal $\Lambda_n$ (cyan curve; bottom, right vertical scale). The inversion pump is defined by equation (\ref{eq:pump}) with a common pump pulse duration $T_{\mathrm{p}}=4$~d at a period of 243.3~d for all features. The length of the systems is set at $L=50$~au, while $T_1=210$~d and $T_2=13.5$~d for OH 1665~MHz, and $T_1=220$~d and $T_2=12.6$~d for methanol 6.7~GHz (both velocities) and 12.2~GHz. Adapted from \citet{Rajabi2023}.}
    \label{fig:G9.62fits}
\end{figure*}
We therefore proceeded to numerically model several flaring features using the MBE in the manner explained in \citet{Rajabi2023}. Results are shown in Figure \ref{fig:G9.62fits} for features in OH 1665~MHz ($v_{\mathrm{lsr}} = +1.7\,\mathrm{km\,s}^{-1}$), methanol 12.2~GHz ($v_{\mathrm{lsr}} = +1.7\,\mathrm{km\,s}^{-1}$) and methanol 6.7~GHz ($v_{\mathrm{lsr}} = +5.0\,\mathrm{km\,s}^{-1}$ and $+8.0\,\mathrm{km\,s}^{-1}$). To do so we fixed the length of the systems to $L=50$~au\footnote{The chosen value of $L=50$~au is to some extent arbitrary since the inversion column density is the relevant parameter for setting the superradiance time-scale (see equation \ref{eq:TR}). That is, any variation in $L$ could be compensated by applying the opposite change in the population inversion density $n$.} and set the excitation from a population inversion pump to
\begin{equation}
\Lambda_{n}\left(z,\tau\right) = \Lambda_{\mathrm{0}} + \sum_{m=0}^\infty \frac{\Lambda_{\mathrm{1},m}}{\cosh^{2}\left[\left(\tau-\tau_0-m\tau_1\right)/T_\mathrm{p}\right]}.\label{eq:pump}
\end{equation}
The constant pump rate $\Lambda_{\mathrm{0}}$, the amplitude $\Lambda_{\mathrm{1},m}$ of pump pulse $m$ at $\tau=\tau_0+m\tau_1$ and the delay $\tau_0$ to line our fit up to the corresponding data are adjusted for all models. Unchanged are the period $\tau_1=243.3$~d and the width of the pump pulse $T_{\mathrm{p}}=4$~d. According to our previous discussion, the choice of the pump pulse's profile is arbitrary. While the relation and de-phasing time-scales are set independently for all transitions, little variations were observed between the different fits. More precisely, we have $T_1=210$~d and $T_2=13.5$~d for OH 1665~MHz, and $T_1=220$~d and $T_2=12.6$~d for methanol 6.7~GHz (both velocities) and 12.2~GHz. 

Despite the small changes in parameters the resulting fits in Figure \ref{fig:G9.62fits} are found to recover the observed time-scales well by simply modulating the amplitude of the pump components $\Lambda_{\mathrm{0}}$ and $\Lambda_{\mathrm{1},m}$. This is because in doing so we are in effect changing the inversion column density $nL$, which in turn directly affects the characteristic time-scale of superradiance $T_\mathrm{R}$ (see equation \ref{eq:TR}). This also explains, for example, why the methanol 6.7~GHz can exhibit significantly varying time-scales at different velocities, as seen for the corresponding $v_{\mathrm{lsr}} = +5.0\,\mathrm{km\,s}^{-1}$ and $+8.0\,\mathrm{km\,s}^{-1}$ features in Figure \ref{fig:G9.62fits}. More models to observations are presented in \citet{Rajabi2023}. Most notable are the cases of unusually shaped flares in methanol 6.7~GHz and OH 1667~MHz; we refer the reader to the corresponding section in that paper. 

Although the flaring periodicity in G9.62+0.20E necessitates an excitation that is different from the case of instantaneous inversion discussed in Sec. \ref{sec:complementarity}, the same type of behaviour is observed. Most notable is the clear disconnect between the duration of the inversion pump pulse (i.e., 4~d) and that of the bursts (e.g., from $\sim40$~d for methanol 6.7~GHz at $v_{\mathrm{lsr}} = +8.0\,\mathrm{km\,s}^{-1}$ to on the order of 100~d for OH 1665~MHz). This is a clear manifestation of a superradiance transient response in the gas hosting the flaring events. As seen in Figure \ref{fig:G9.62fits}, while the interaction of the fast and brief pump pulses drives a rapid increase in the population inversion level and subsequently stimulates a transient superradiance response in the system, the duration of the superradiance burst bears no relation to that of the excitation. We also note that adopting a fixed value for the duration of the pump pulses is a reasonable assumption, as it is likely that a single source is at the root of the 243~d periodicity of the flares. Time-scales intrinsic to the pump (i.e., the period and pulse duration) should be the same for all transition/species used for the monitoring. On the other hand, the coupling of the inversion pump signal is bound to vary with position within G9.62+0.20E and wavelength. The dependence of the superradiance response on the strength of the excitation and the parameters characterizing the molecular transition/species used to probe the regions under study renders it an attractive and powerful mechanism to make sense of the diverse flaring time-scales and profiles observed.  

\begin{figure}
    \centering
    \includegraphics[width=0.6\textwidth]{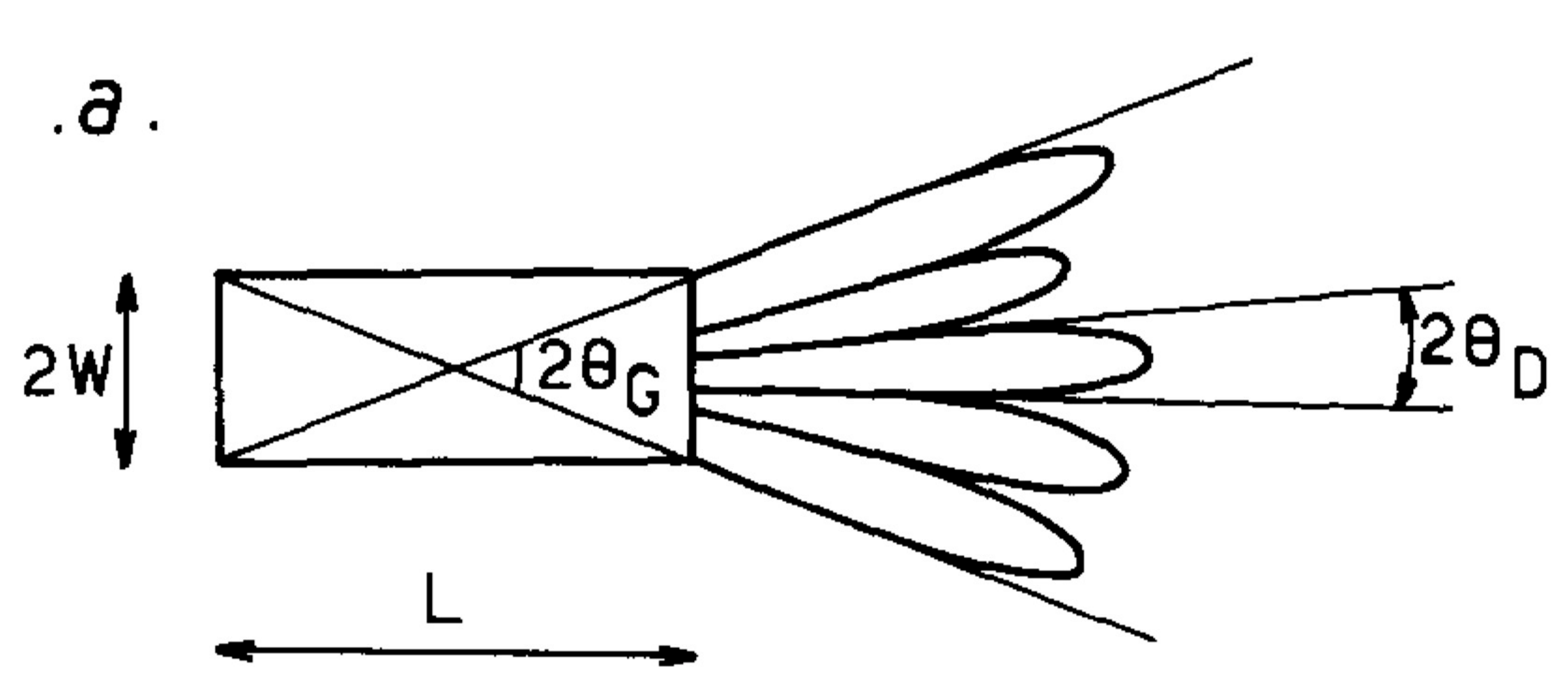}
    \caption{Schematic for a region of length $L$ and a cross-section of width $2W$ exceeding that of a single sample, as defined in Sec. \ref{sec:complementarity}. This region has a Fresnel number greater than unity and supports a geometrical angle $\theta_\mathrm{G}=W/L$ at its end-fire. It will therefore contain several independent samples of diffraction angle $\theta_\mathrm{D}=\lambda/W$ during a superradiance event. Although radiating coherently individually, these samples combine in a non-coherent manner to make up the total intensity measured at the output. For an astronomical maser-hosting source, this intensity is likely to appear non-coherent to an observer. Taken from \citet{Gross1982}.}
    \label{fig:Fresnel}
\end{figure}
Finally, we end with a word concerning the potential effects that the coherent nature of superradiance might be suspected to have on the characteristics of observations such as those presented in this section. That is, we have already stated that superradiance bursts scale in intensity with the square of the number of molecules involved in the process and inversely with that number in duration \citep{Dicke1954,Feld1980,Gross1982,Rajabi2020}. However, one should not necessarily expect coherence to be readily detected in astronomical sources hosting flares such as those analyzed here. The reason for this is presented in Figure \ref{fig:Fresnel} and rests on the relative size of a coherent superradiance sample and an astronomical maser-hosting region. Considering as an example the OH~1665~MHz sample resulting from the fit presented in Figure \ref{fig:G9.62fits}, we find that its radius is limited to $\sim 10^{-5}$~au to ensure a Fresnel number of unity. Evidently, this size is several orders of magnitude smaller than the spot size of a maser-hosting region. For example, a region of 10~au undergoing a flaring event would necessarily break up in a very large number of independent and uncorrelated samples, in the manner shown in Figure \ref{fig:Fresnel}. Although individually radiating coherently, these samples combine in a non-coherent manner to make up the total intensity measured at the output. For an astronomical maser-hosting source, this intensity is thus likely to appear non-coherent to an observer. Our superradiance OH 1665~MHz sample of Figure \ref{fig:g9.62-all} outputs a peak flux density of $\approx 5\times 10^{-36}$~erg~s$^{-1}$~cm$^{-2}$ at 5.2~kpc (or $\sim 10^{-5}$~Jy in the bandwidth of $\sim 10^{-7}$~Hz approximately associated to the duration of a flare). Only a small fraction of a typical maser-hosting is therefore needed to account for the detected radiation intensities.      

\section*{Acknowledgements}
M.H.'s research is funded through the Natural Sciences and Engineering Research Council of Canada Discovery Grant RGPIN-2016-04460. The 6.7 GHz methanol maser data obtained by the Hitachi 32-m telescope is a part of the Ibaraki 6.7 GHz Methanol Maser Monitor (iMet) program. The iMet program is partially supported by the Inter-university collaborative project `Japanese VLBI Network (JVN)’ of NAOJ and JSPS KAKENHI grant no. JP24340034, JP21H01120, and JP21H00032 (YY).

\bibliographystyle{iaulike}
\bibliography{scibib}

\end{document}